\documentclass[aps,prd,twocolumn,superscriptaddress,nofootinbib,showpacs,10pt]{revtex4-1}
\usepackage[utf8]{inputenc}
\usepackage[english]{babel}
\usepackage{amsmath}
\usepackage{amsfonts}
\usepackage{amssymb}
\usepackage{graphics}
\usepackage{graphicx}
\usepackage[left=1.8cm,right=1.8cm,top=2.0cm,bottom=2.0cm]{geometry}
\usepackage[usenames,dvipsnames,svgnames]{xcolor}  
\usepackage{subfigure}
\usepackage{hyperref}   
\hypersetup{colorlinks=true,linkcolor=beamer@PRD, citecolor=beamer@PRD}
\definecolor{beamer@PRD}{RGB}{46,48,146}
\newcommand\myref[1]{\textcolor{beamer@PRD}{(}\ref{#1}\textcolor{beamer@PRD}{)}}

\usepackage{todonotes}
\usepackage{braket}
\begin{document}
\title{Generalized photon-subtracted squeezed vacuum states}
\author{Sanjib Dey}\email{dey@iisermohali.ac.in}
\affiliation{Department of Physical Sciences, Indian Institute of Science Education and  Research Mohali, \protect\\ Sector 81, SAS Nagar, Manauli 140306, India}
\author{Sarika S. Nair}
\affiliation{Department of Physical Sciences, Indian Institute of Science Education and  Research Mohali, \protect\\ Sector 81, SAS Nagar, Manauli 140306, India}
\begin{abstract}
We construct a generalized version of the photon-subtracted squeezed vacuum states (PSSVS), which can be utilized to construct the same for nonlinear, deformed and any usual quantum mechanical models beyond the harmonic oscillator. We apply our general framework to trigonometric P\"oschl-Teller potential and show that our method works accurately and produces a proper nonclassical state.  We analyze the nonclassicality of the state using three different approaches, namely, quadrature squeezing, photon number squeezing and Wigner function and indicate how the standard definitions of those three techniques can be generalized and utilized to examine the nonclassicality of any generalized quantum optical states including the PSSVS. We observe that the generalized PSSVS are always more nonclassical than those arising from the harmonic oscillator. Moreover, within some quantification schemes, we find that the nonclassicality of the PSSVS increases almost proportionally with the number of photons subtracted from the generalized squeezed vacuum state. Thus, generalized PSSVS may provide an additional freedom with which one can regulate the nonclassicality and obtain an appropriate nonclassical state as per requirement.    
\end{abstract}


\maketitle
\section{Introduction} \label{sec1}
Wigner function is an excellent framework for the study of quantum optical states. States with Gaussian Wigner function have certain applications in quantum information processing and they have been studied extensively in the literature. More recent studies show that the states with non-Gaussian Wigner function are also very important in the field, especially those having a negative Wigner function. Negativity in Wigner function is a sufficient condition for nonclassicality and the associated states are extremely useful in entanglement distillation \cite{Eisert_Scheel_Plenio}, quantum computing \cite{Bartlett_Sanders}, etc. Several important non-Gaussian states have been studied and are found to exist in real life. For instance, in \cite{Wenger_Tualle-Brouri_Grangier}, the authors revealed an experimental method for preparing non-Gaussian states  from a single mode squeezed light using the homodyne detection technique. A similar technique was utilized to detect a single-photon state in \cite{Lvovsky_etal}. Alongside several theoretical studies to detect two-photon \cite{Chen_etal1,Chen_etal2} and three-photon \cite{Chen_etal} nonclassical states, there are plenty of experimental results for the preparation of various nonclassical states including Schr\"odinger cat states \cite{Ourjoumtsev_Jeong_Tualle-Brouri_Grangier}, squeezed states \cite{Wu_Kimble_Hall_Wu}, photon-subtracted squeezed/squeezed vacuum states \cite{Wakui_Takahashi_Furusawa_Sasaki}, etc.

While standard quantum optical non-Gaussian states have immense importance, a lot of recent studies indicate that the generalized quantum optical states play an important role in quantum information processing, see; for instance \cite{Dodonov,Dey_Fring_Hussin_Review} for some reviews on the development. Generalization to quantum optics can be performed in various ways, some of the well-known approaches include the nonlinear generalization \cite{Manko_Marmo_Sudarshan_Zaccaria,Sivakumar,Filho_Vogel}, $q$-deformation \cite{Biedenharn,Macfarlane}, etc. A common goal in all such frameworks is to construct the quantum optical states for other quantum mechanical potentials apart from the harmonic oscillator. One of the notable usefulnesses of generalizing the quantum optical models is that it brings in additional degrees of freedom to the system by which one can improve several crucial properties of the system \cite{Dey_Hussin_atom}. More importantly, it has been shown that generalized nonclassical states provide higher degree of nonclassicality compared to the usual nonclassical states \cite{Dey,Dey_Hussin_Photon,Zelaya_Dey_Hussin}, which can be exploited to enhance the quantum entanglement of various nonclassical states within some protocols  \cite{Kim_Son_Buzek_Knight,Dey_Hussin,Dey_Fring_Hussin}. The usage of the generalized quantum optical states is not limited to the theoretical studies but there have been ample experimental investigations based on Kerr type nonlinear cavities \cite{Wang_Goorskey_Xiao,Gambetta_etal,Yan_Zhu_Li}.

In this article, we construct a class of non-Gaussian nonclassical states, namely the photon-subtracted squeezed vacuum sates (PSSVS) arising from the generalized framework. We provide a generic analytical prototype which can be utilized to construct generalized PSSVS for any nonlinear, deformed and quantum mechanical models. As per the definition, the states are bound to exhibit nonclassical properties, which we verify using several independent methods, namely, quadrature squeezing, photon distribution function and negativity in Wigner function. One of the striking features of such states is that the nonclassicality increases proportionally with the number of photons being subtracted from the states, which is visible evidently in two of our approaches, the number squeezing and the Wigner function. Therefore, such states may be employed methodically to produce more nonclassicality compared to the generalized squeezed vacuum states. We also notice that the generalized PSSVS are always more nonclassical than those emerging from the harmonic oscillator potential, and such a phenomenon occurs consistently in all three given methods. Thus, we believe that such states will be a compelling successor for quantum information theories in many aspects.   

In Sec.\,\ref{sec2}, we describe the methodology for the construction of generalized PSSVS along with a complete analytical description of the states. Sec.\,\ref{sec3} is composed of the solution of the P\"oschl-Teller potential and a contemporary procedure for extracting the required information for the construction of the PSSVS for a particular generalized potential. We, then, provide a detailed analysis of various nonclassical features of the PSSVS of the P\"oschl-Teller potential in Sec.\,\ref{sec4}. Finally, our conclusions are stated in Sec.\,\ref{sec5}.
\section{Generalized photon-subtracted squeezed vacuum}\label{sec2}
PSSVS can be constructed by subtracting single or multiple number of photons from the squeezed vacuum states \cite{Agarwal_Biswas_1}. Before moving to the full fledged construction of generalized PSSVS, let us commence with the discussion of the method of construction of the harmonic oscillator squeezed vacuum states $|\zeta\rangle$. Usually, it follows from the operation of the squeezing operator $\hat{S}(\zeta)=\exp{[(\zeta^\ast \hat{a}^2-\zeta \hat{a}^{\dagger^2})/2]}$ on the vacuum state $|0\rangle$ and the result is entirely equivalent to that obtained from the following definition \cite{Gerry_Knight_Book}
\begin{equation}\label{SVS}
(\hat{a}\mu+\hat{a}^\dagger\lambda)|\zeta\rangle =0,
\end{equation}  
with $\mu=\cosh r$ and $\lambda=e^{i\theta}\sinh r$. Here, the squeezing parameter $\zeta$ is represented in the polar form $\zeta=re^{i\theta}$. Note that the relation \myref{SVS} does not originate from an independent source, rather, it is a byproduct of the first definition $|\zeta\rangle=\hat{S}(\zeta)|0\rangle$. It is worth mentioning that several mixed terminologies are used for such states in the literature, for example, in \cite{Agarwal_Biswas_1}, the authors claim that they study the squeezed states, however, originally they have studied the squeezed vacuum states. We stay clear of this confusion by specifying the states as follows. We attribute the name `squeezed vacuum states' to the states that are generated by the action of the squeezing operator $\hat{S}(\zeta)$ on the vacuum state $|0\rangle$, i.e.\,$|\zeta\rangle=\hat{S}(\zeta)|0\rangle$. However, when we refer the `squeezed states', we mean the states originating from the operation of the squeezing operator on the coherent states $|\alpha\rangle$ instead, viz.\,$|\zeta,\alpha\rangle=\hat{S}(\zeta)|\alpha\rangle$. In this article, we study the generalization of squeezed vacuum states $|\zeta\rangle$. Nevertheless, the first step towards the generalization is to find a general set of ladder operators $\hat{A}\equiv \hat{a}f(\hat{n})=f(\hat{n}+1)a,~\hat{A}^\dagger\equiv f(\hat{n})\hat{a}^\dagger=\hat{a}^\dagger f(\hat{n}+1)$, whose action on the Fock states can be realized as follows
\begin{eqnarray}\label{GenLad}
\hat{A}|n\rangle &=& \sqrt{n}f(n)|n-1\rangle, \\
\hat{A}^\dagger|n\rangle &=& \sqrt{n+1}f(n+1)|n+1\rangle, \notag 
\end{eqnarray}
so that the number operator of the generalized system can be considered as $\hat{A}^\dagger \hat{A}\equiv \hat{n}f^2(\hat{n})$. Here $f(\hat{n})$ is an operator valued function of the harmonic oscillator number operator $\hat{n}\equiv \hat{a}^\dagger \hat{a}$ and, it is an entirely general function. The exact expression of $f(\hat{n})$ can be extracted from the knowledge of the eigenvalues of the corresponding Hamiltonian assuming that it can be factorized in terms of the generalized number operator $\hat{A}^\dagger \hat{A}$. Any constant terms that may appear in the Hamiltonian can be realized by a proper rescaling of the eigenvalues of the composite system containing $\hat{A}$ and $\hat{A}^\dagger$. Thus, the knowledge of the eigenvalues of any quantum mechanical Hamiltonian will ensure the explicit form of the function $f(n)$. The method is usually familiar as the nonlinear generalization \cite{Manko_Marmo_Sudarshan_Zaccaria,Sivakumar,Filho_Vogel} and the endeavor has been accepted widely; see, for instance  \cite{Trifonov,Quesne_Penson_Thachuk,Kwek_Kiang,Naderi_Soltanolkotabi_Roknizadeh,Ching_Ng,Ramirez_Reboiro, Fakhri_Hashemi}. However, a direct replacement of the generalized ladder operators \myref{GenLad} in \myref{SVS} does not necessarily yield the generalized PSSVS, rather, the resulting state is some other state whose name is not known to us. Indeed there exist some studies where the generalized ladder operators \myref{GenLad} have been used directly in \myref{SVS}, see; for instance \cite{Noormandipour_Tavassoly}, however, the results following such an approach may be regarded incorrect. Notice that the relation $(\hat{A}\mu+\hat{A}^\dagger\lambda)|\zeta,f\rangle=0$ does not originate from the original definition of generalized squeezed vacuum states, i.e.\,$|\zeta,f\rangle=\hat{S}(\zeta,f)|0\rangle$. This is because the generalized ladder operators satisfy the commutation relation  
\begin{equation}\label{AAdag}
[\hat{A},\hat{A}^\dagger]=(\hat{n}+1)f^2(\hat{n}+1)-\hat{n}f^2(\hat{n}),
\end{equation}
and, thus, the generalized squeezing operator $\hat{S}(\zeta,f)=\exp{[(\zeta^\ast \hat{A}^2-\zeta \hat{A}^{\dagger^2})/2]}$ can no longer be disentangled. Therefore, it is impossible to reach to the definition $(\hat{A}\mu+\hat{A}^\dagger\lambda)|\zeta,f\rangle=0$ from the original definition $|\zeta,f\rangle=\hat{S}(\zeta,f)|0\rangle$. To overcome this problem, let us introduce a set of auxiliary ladder operators $\hat{B}=\hat{a}[1/f(\hat{n})]=[1/f(\hat{n}+1)]a$ and $\hat{B}^\dagger=[1/f(\hat{n})]\hat{a}^\dagger=\hat{a}^\dagger [1/f(\hat{n}+1)]$ resulting to a new set of commutation relations $[\hat{A},\hat{B}^\dagger]=[\hat{B},\hat{A}^\dagger]=1$. Therefore, one can consider a new set of conjugate ladder operators $\hat{A},\hat{B}^\dagger$ or $\hat{B},\hat{A}^\dagger$ which will allow the disentanglement of the squeezing operator. The method was introduced in \cite{Roy_Roy} in order to explore a new type of coherent states which become compatible with both of the definitions of nonlinear coherent states, $|\alpha,f\rangle=\hat{D}(\alpha,f)|0\rangle$ and $A|\alpha,f\rangle=\alpha|\alpha,f\rangle$, with $D(\alpha,f)=\exp[\alpha\hat{A}^\dagger-\alpha^\ast\hat{A}]$ being the optical displacement operator. 

Let us now describe an appropriate way to construct the generalized squeezed vacuum states. We define the squeezed vacuum as
\begin{equation}\label{GSVS1}
|\zeta,f\rangle=\hat{S}(\zeta,f)|0\rangle=e^{\frac{1}{2}(\zeta^\ast \hat{A}^2-\zeta \hat{B}^{\dagger^2})}|0\rangle,
\end{equation}
which leads to the alternative definition
\begin{equation}\label{GSVS2}
(\hat{A}\mu+\hat{B}^\dagger\lambda)|\zeta,f\rangle=0,
\end{equation}
with $\mu=\cosh r$ and $\lambda=e^{i\theta}\sinh r$. It is straightforward to check that the relation \myref{GSVS1} leads to \myref{GSVS2}. Considering the state $|\zeta,f\rangle$ to be residing in the photon number space $|n\rangle$, we can decompose the state in the Fock basis, i.\,e.\,$|\zeta,f\rangle=\sum_{n=0}^{\infty}C_{n,f}|n\rangle$. Thereafter, using this expression in \myref{GSVS2}, we obtain the recursion relation for the expansion coefficients
\begin{equation}\label{recursion}
C_{m+1,f}=-\frac{e^{i\theta}\tanh r}{f(m)f(m+1)}\sqrt{\frac{m}{m+1}}C_{m-1,f},
\end{equation}
which when solved in the even basis, we obtain the required state as follows
\begin{equation}\label{GSVS}
|\zeta,f\rangle=\frac{1}{\mathcal{N}_{\zeta,f}}\displaystyle\sum_{n=0}^{\infty}(-1)^n\frac{e^{in\theta}(\tanh r)^n\sqrt{(2n)!}}{2^nn!f(2n)!}|2n\rangle.
\end{equation}
The normalization constant is given by
\begin{equation}
\mathcal{N}^2_{\zeta,f}=\displaystyle\sum_{n=0}^{\infty}\frac{(2n)!(\tanh r)^{2n}}{4^n(n!)^2[f(2n)!]^2}.
\end{equation}
Note that the solution of \myref{recursion} in the odd basis will lead to the generalized squeezed first excited state and we are not interested in it. We shall rather explore the effects of the photon subtraction from the generalized squeezed vacuum states \myref{GSVS}.

The generalized PSSVS can, in principle, be constructed by subtracting $m$ number of photons from the squeezed vacuum states  \myref{GSVS}, i.e. by operating $\hat{A}^m$ on the state $|\zeta,f\rangle$. However, subtraction of arbitrary number of photons will impose a restriction on the state that $2n-m$ has to be positive. We can avoid such a restriction by studying the states in even and odd bases separately and the closed forms of the generalized PSSVS are given by
\begin{eqnarray} \label{EPSSVS}
&& |\zeta,f,m\rangle_\text{e} = \hat{A}^{2m}|\zeta,f\rangle \\
&& = \frac{1}{\mathcal{N}^\text{e}_{\zeta,f,m}}\displaystyle\sum_{n=0}^{\infty}\frac{(-\tanh r)^{k}e^{ik\theta}(2k)!}{2^{k}k!\sqrt{(2n)!}f(2n)!}|2n\rangle, \notag 
\end{eqnarray}  
and
\begin{eqnarray} \label{OPSSVS}
&& |\zeta,f,m\rangle_\text{o} = \hat{A}^{2m+1}|\zeta,f\rangle \\
&&  = \frac{1}{\mathcal{N}^\text{o}_{\zeta,f,m}}\displaystyle\sum_{n=0}^{\infty}\frac{(-\tanh r)^{k'}e^{ik'\theta}(2k')!}{2^{k'}k'!\sqrt{(2n+1)!}f(2n+1)!}|2n+1\rangle, \notag
\end{eqnarray}
with the normalization constants being
\begin{eqnarray}
&& \left[\mathcal{N}^\text{e}_{\zeta,f,m}\right]^2=\displaystyle\sum_{n=0}^{\infty}\frac{(\tanh r)^{2k}[(2k)!]^2}{4^k(k!)^2(2n)![f(2n)!]^2}, 
\\
&& \left[\mathcal{N}^\text{o}_{\zeta,f,m}\right]^2=\displaystyle\sum_{n=0}^{\infty}\frac{(\tanh r)^{2k'}[(2k')!]^2}{4^{k'}(k'!)^2(2n+1)![f(2n+1)!]^2},
\end{eqnarray}
where $k=m+n$ and $k'=k+1$. Here $m$ is absolutely arbitrary and replacement of $m=0$ in \myref{EPSSVS} and \myref{OPSSVS} yield the squeezed vacuum states and single PSSVS, respectively.
\begin{figure*}
  \subfigure[]{
\includegraphics[width=5.7cm]{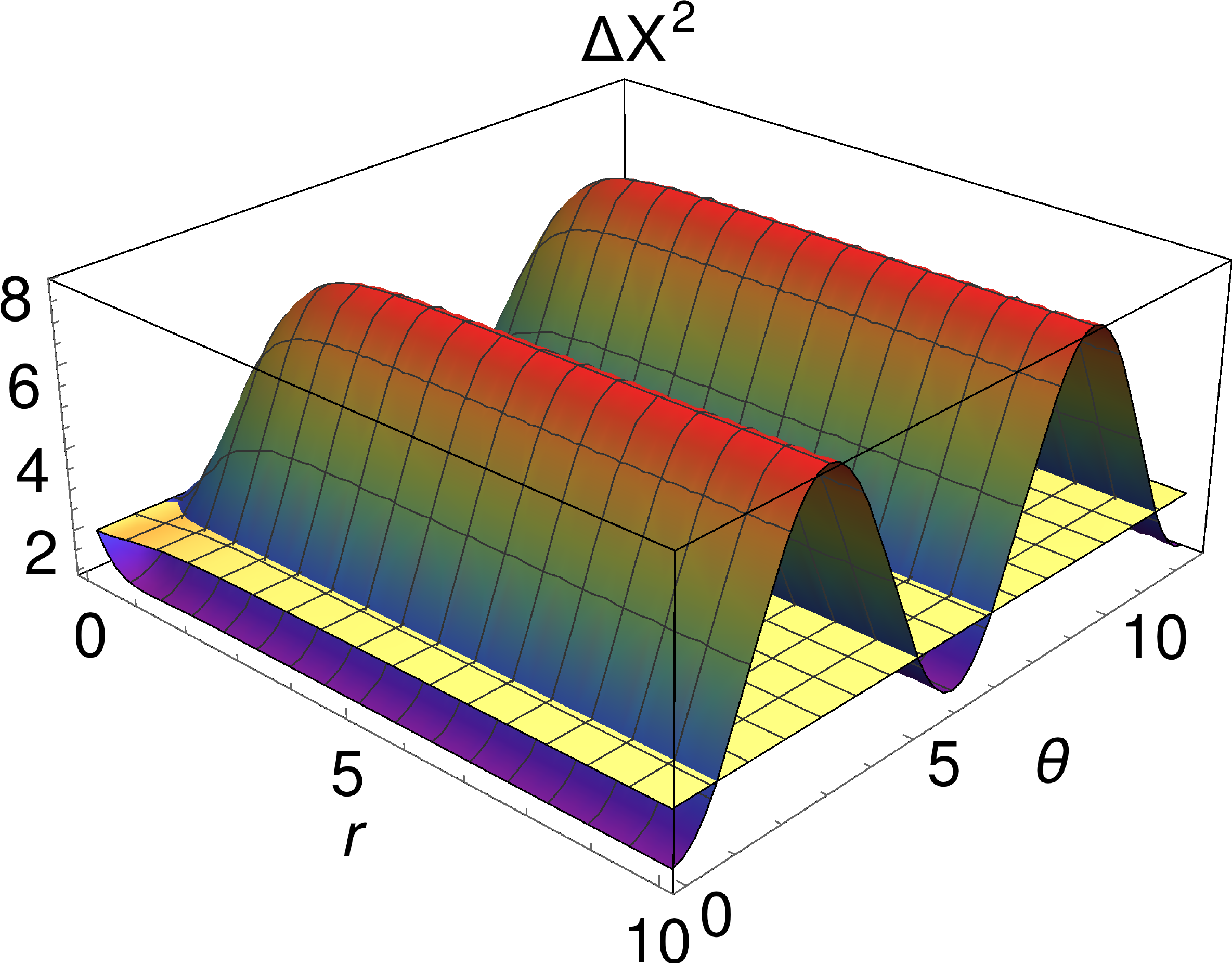}
\label{fig3a}}
  \subfigure[]{
    \includegraphics[width=5.7cm]{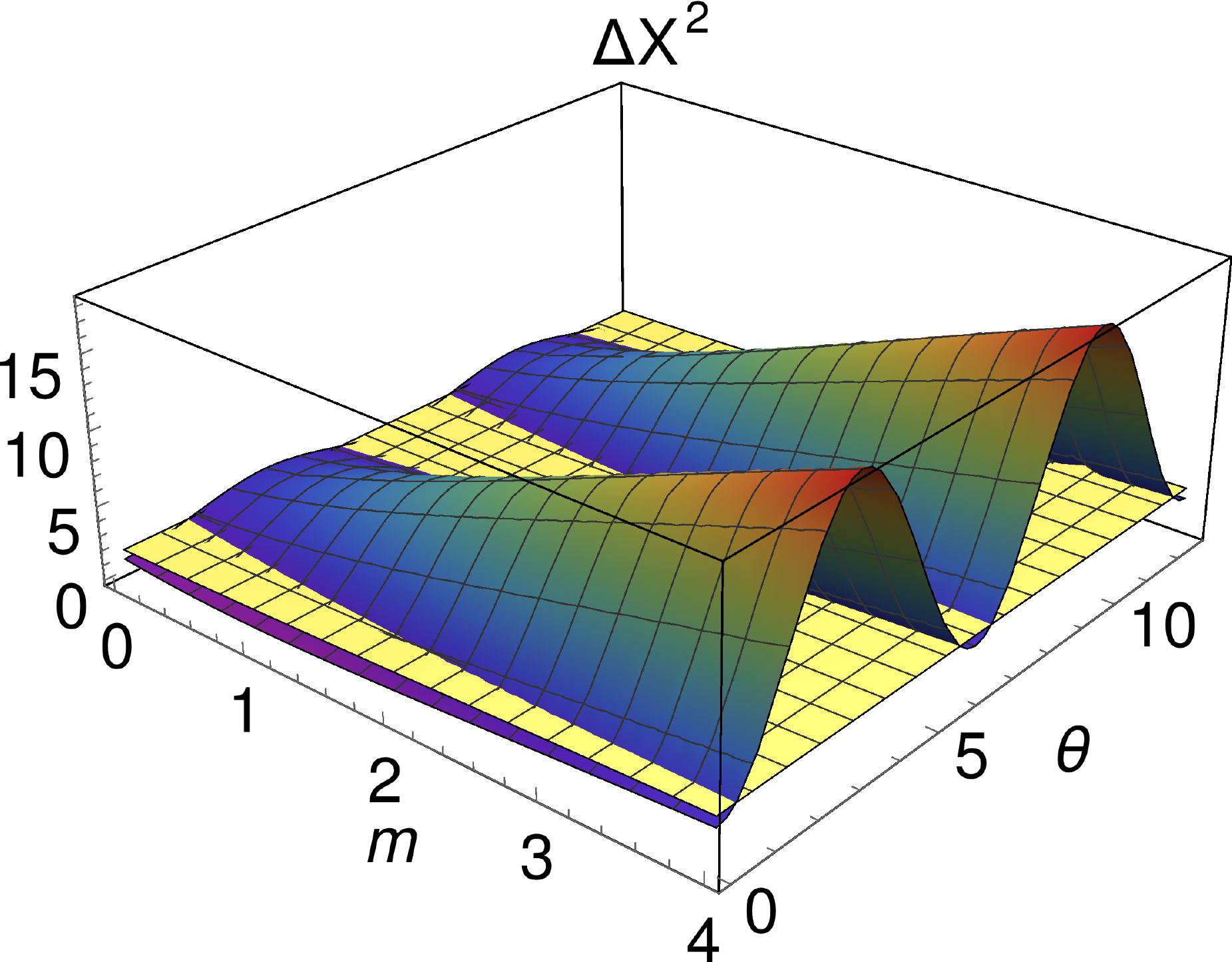}
    \label{fig3b}} 
\subfigure[]{
  \includegraphics[width=5.7cm]{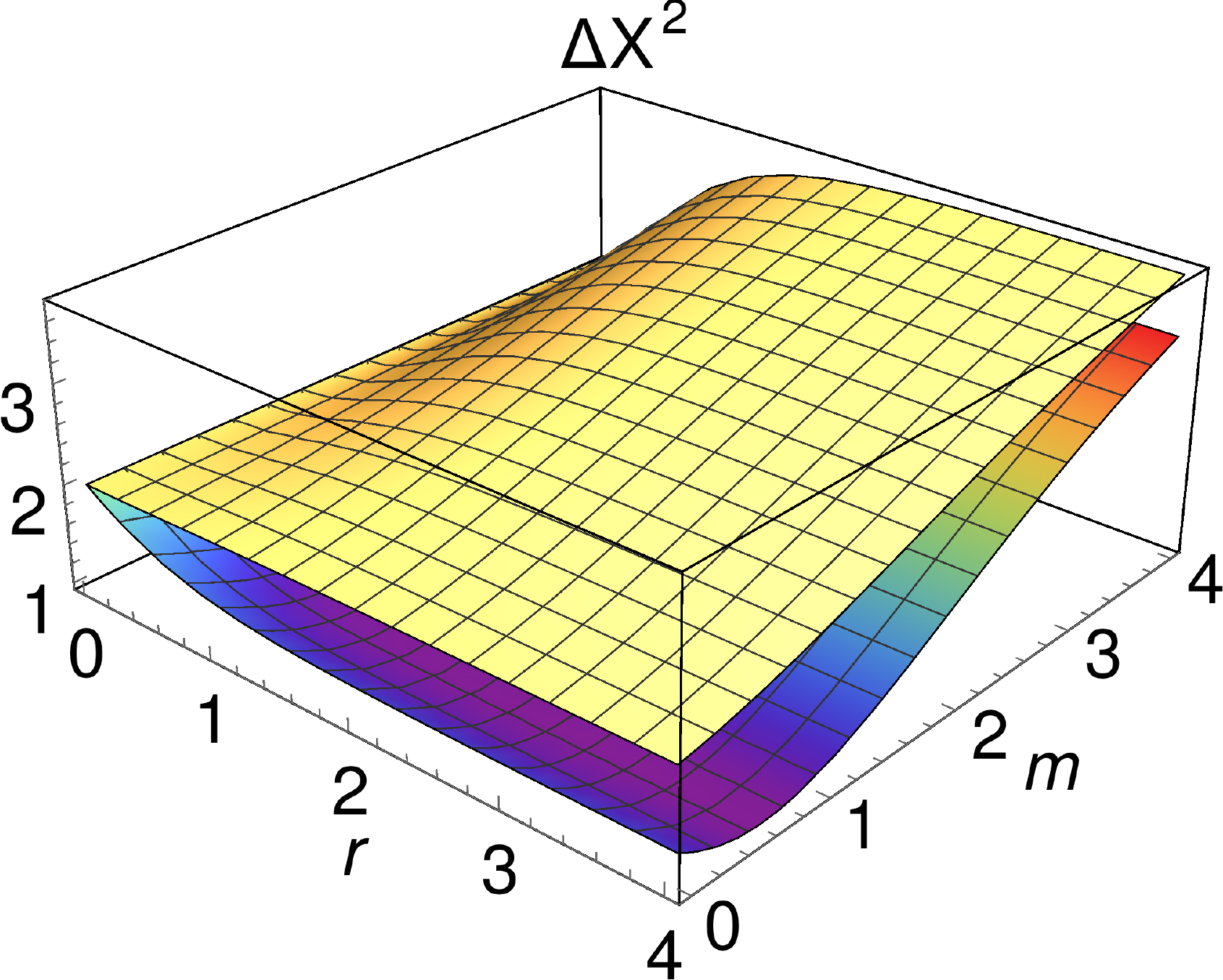}
  \label{fig3c}}
\subfigure[]{
  \includegraphics[width=5.7cm]{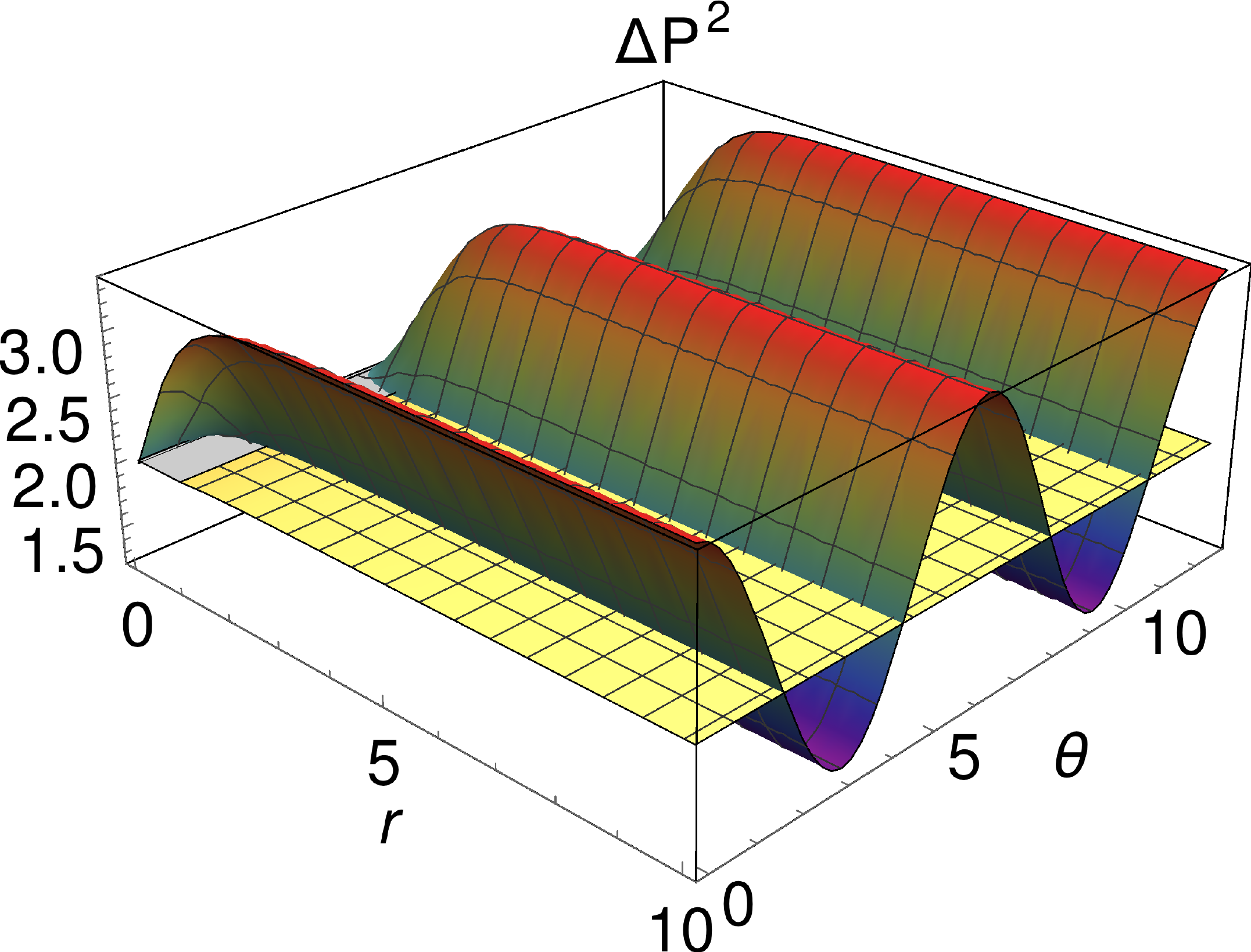}
  \label{fig3d}}
\subfigure[]{
    \includegraphics[width=5.7cm]{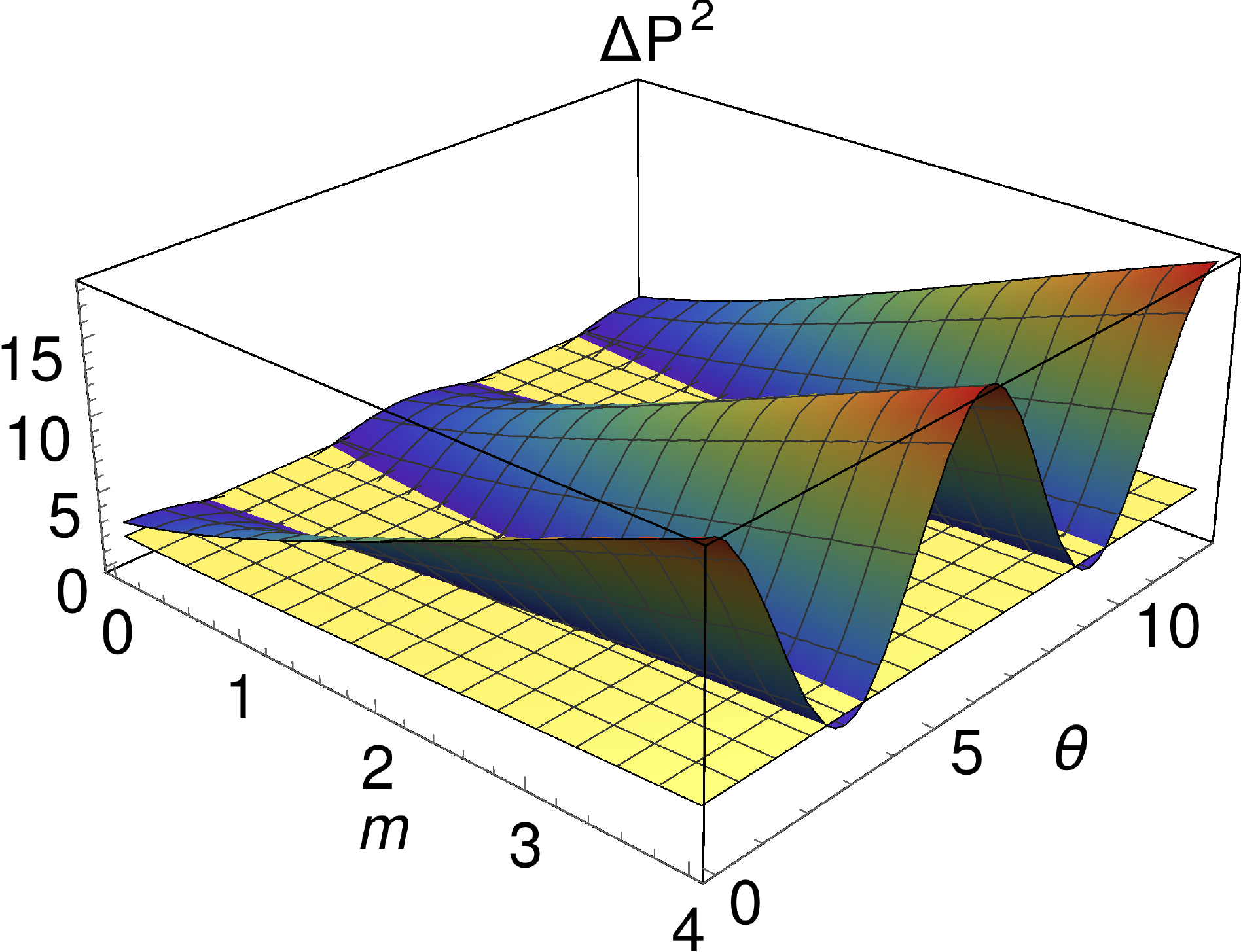}
    \label{fig3e}} 
\subfigure[]{
  \includegraphics[width=5.7cm]{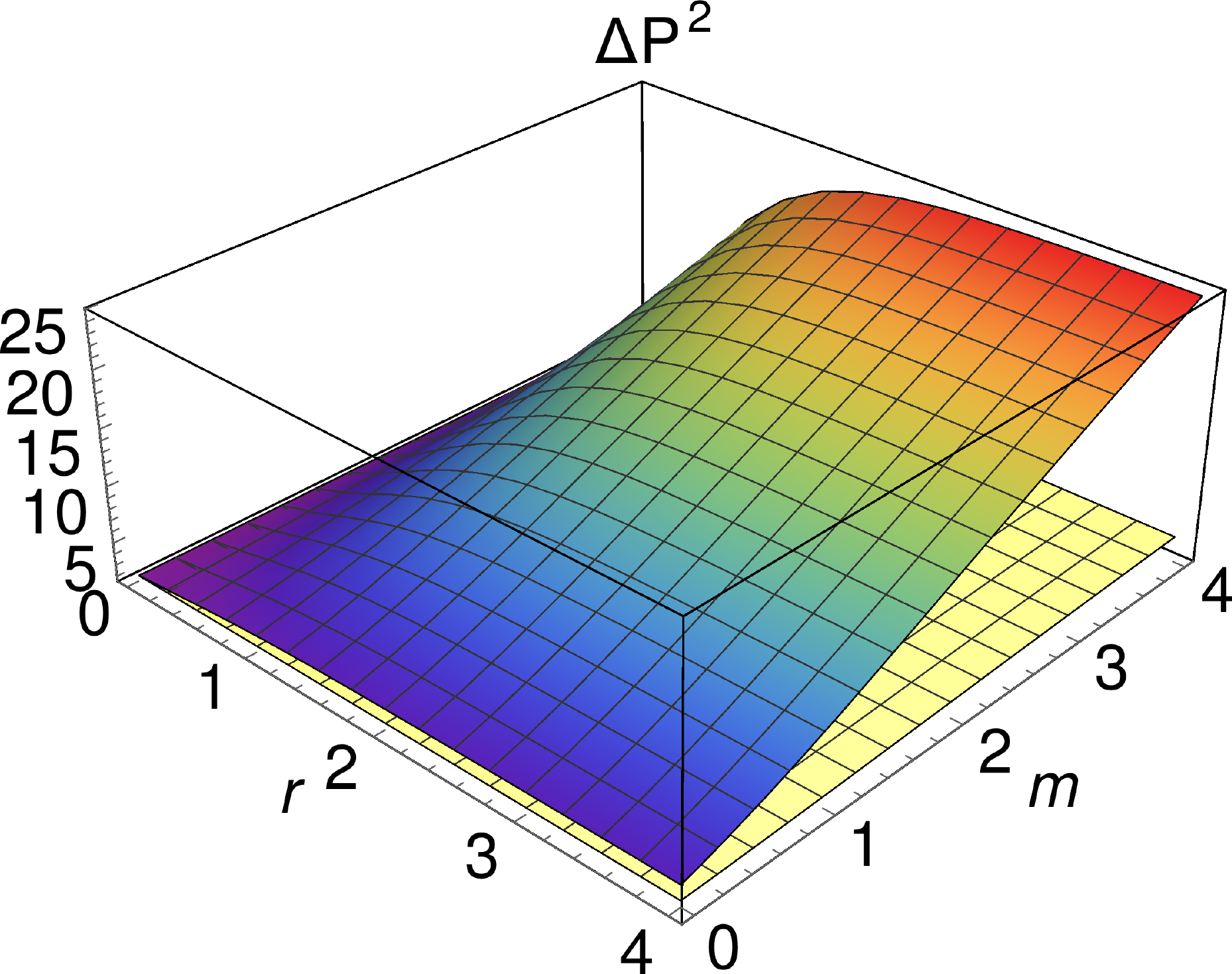}
  \label{fig3f}}
 \caption{(Color online) Square of the uncertainties of $X$ and $P$ quadratures for even generalized PSSVS for $m=1$ in \subref{fig3a} $\&$ \subref{fig3d}, $r=1$ in \subref{fig3b} $\&$ \subref{fig3e} and $\theta = 0$ in \subref{fig3c} $\&$ \subref{fig3f}. In all the plots $\lambda$ and $\kappa$ are chosen to be $1.5$. The yellow surfaces in all the figures show the variation of the RHS of the generalized Robertson uncertainty relation.}
  \label{QS_even}
\end{figure*}
\begin{figure*}
  \subfigure[]{
\includegraphics[width=5.7cm]{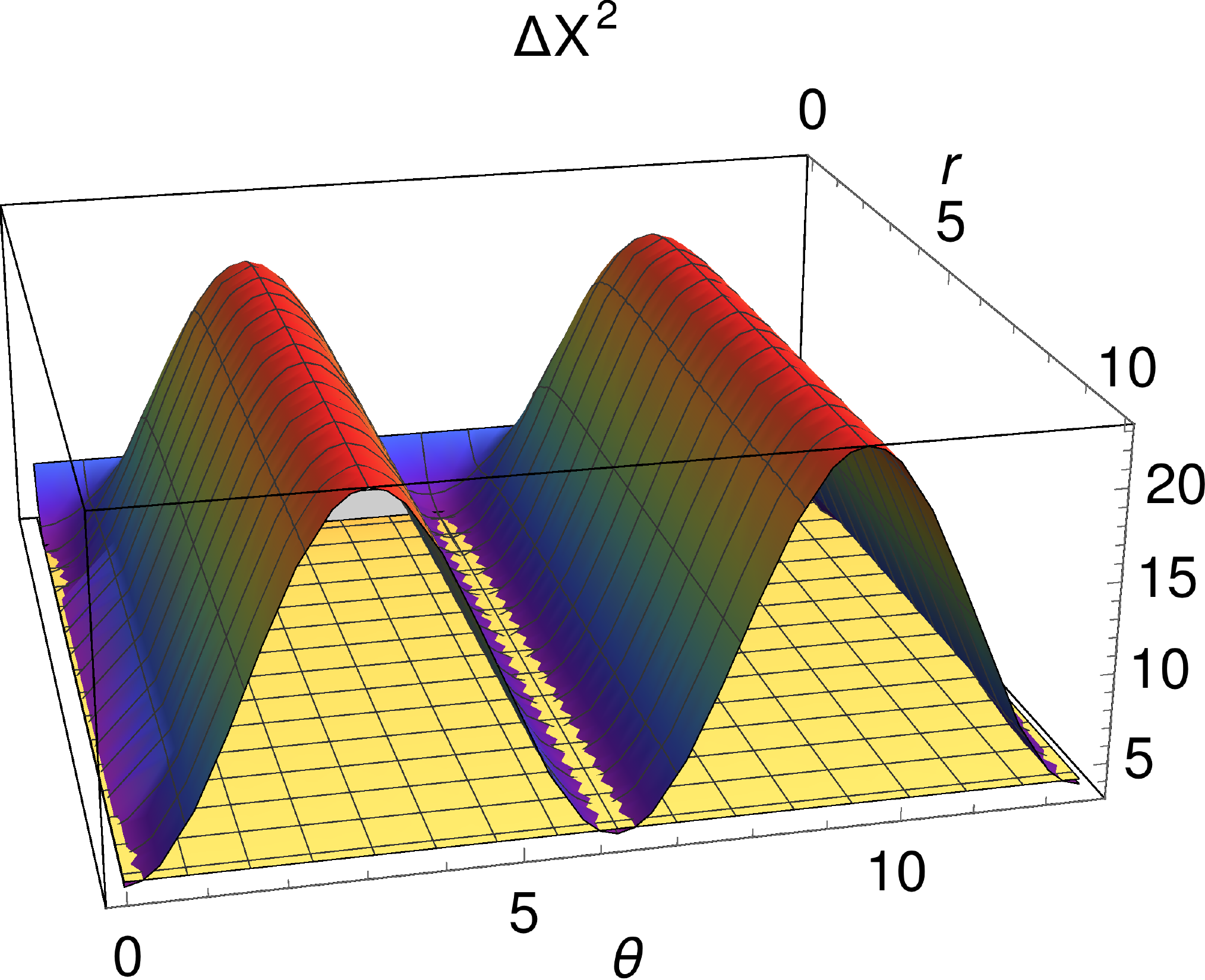}
\label{fig4a}}
  \subfigure[]{
    \includegraphics[width=5.7cm]{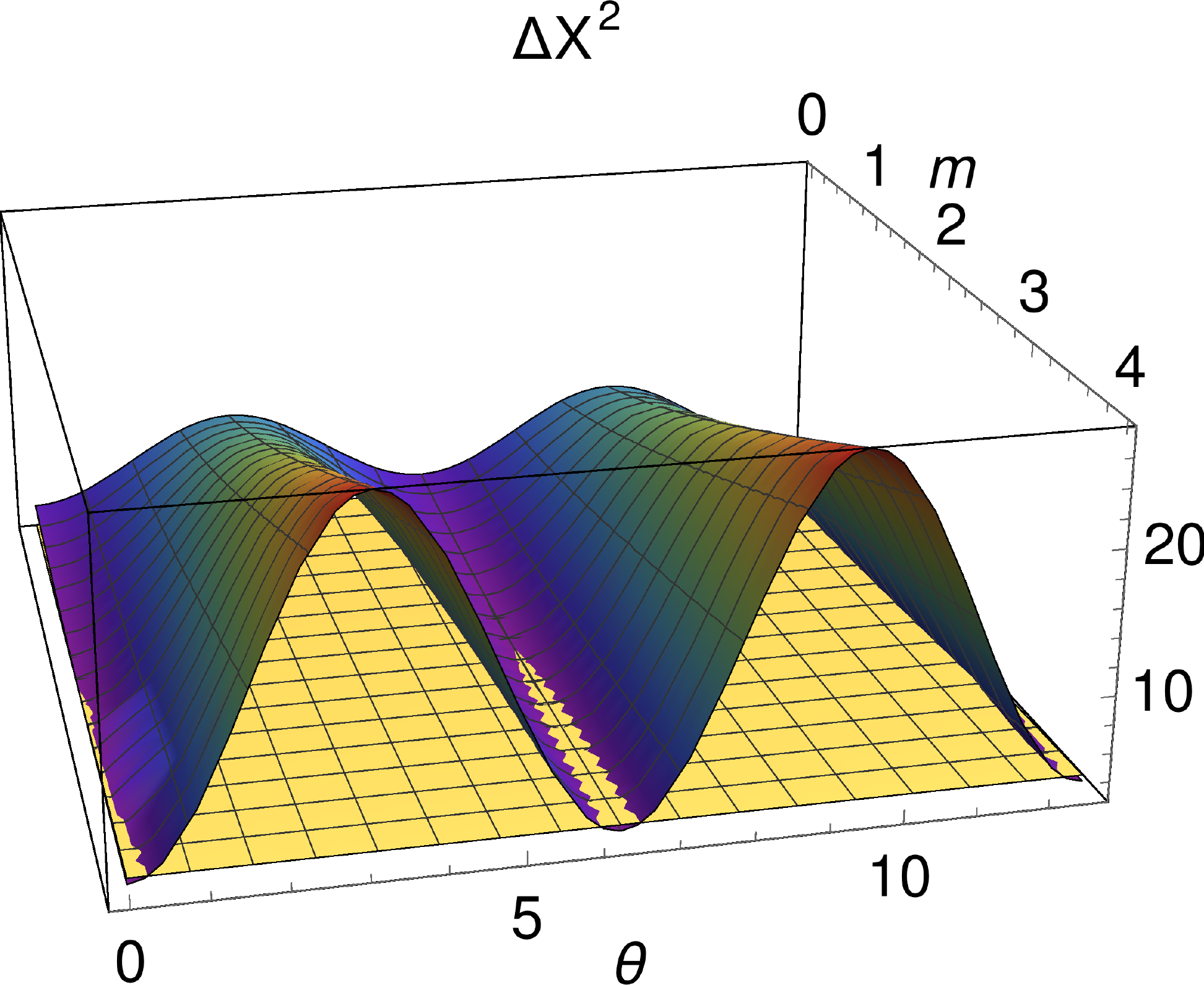}
    \label{fig4b}} 
\subfigure[]{
  \includegraphics[width=5.7cm]{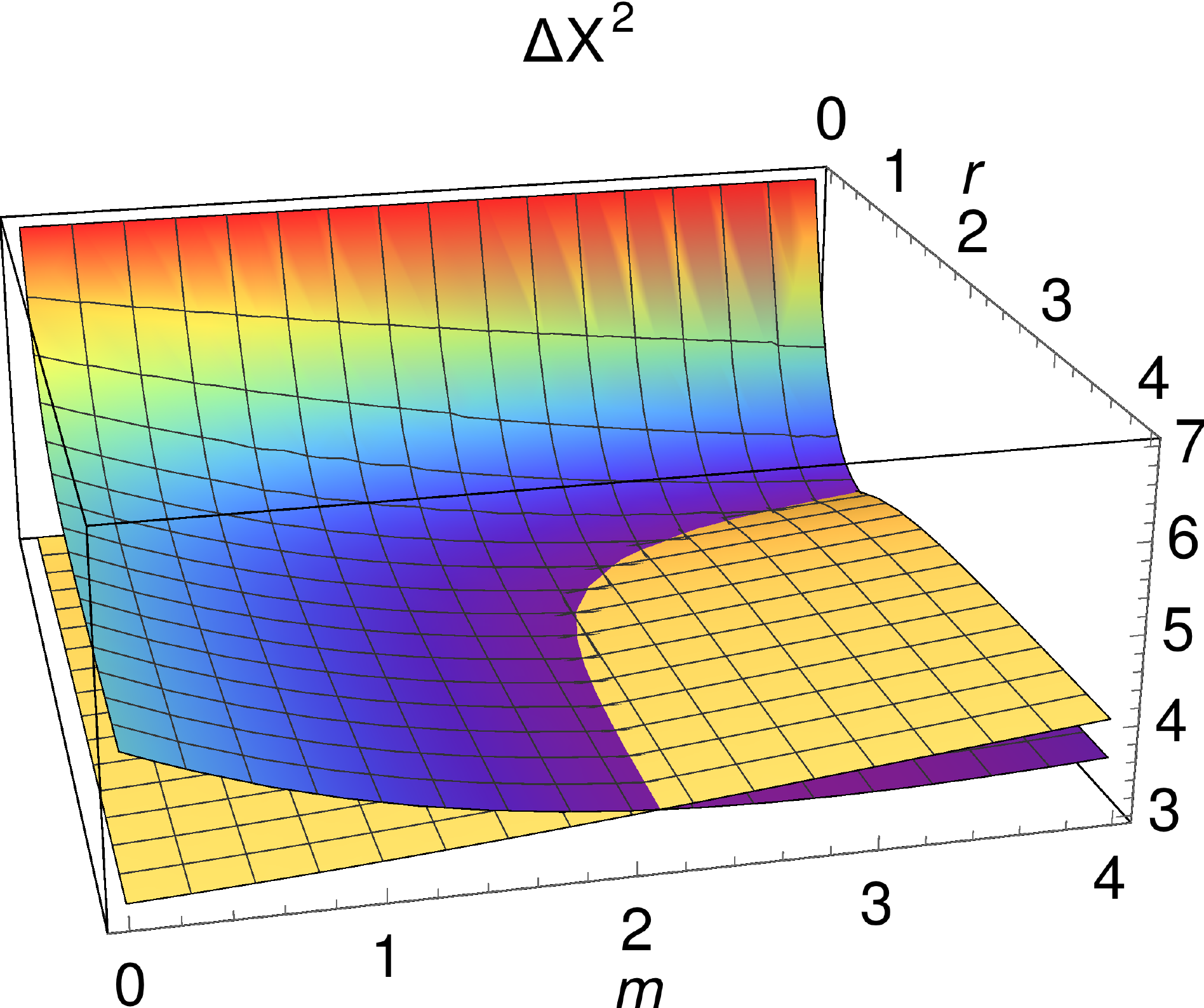}
  \label{fig4c}}
\subfigure[]{
  \includegraphics[width=5.7cm]{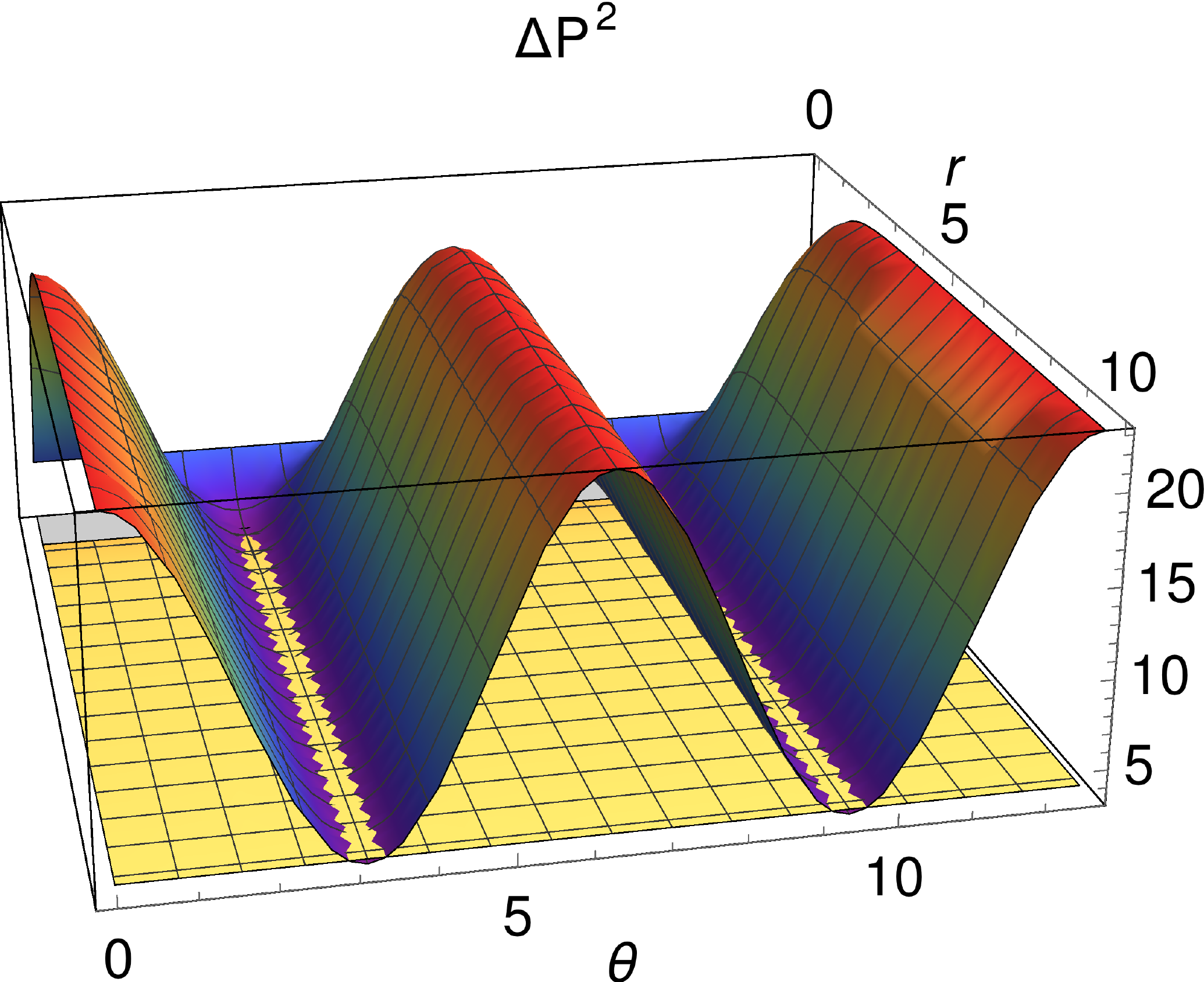}
  \label{fig4d}}
\subfigure[]{
    \includegraphics[width=5.7cm]{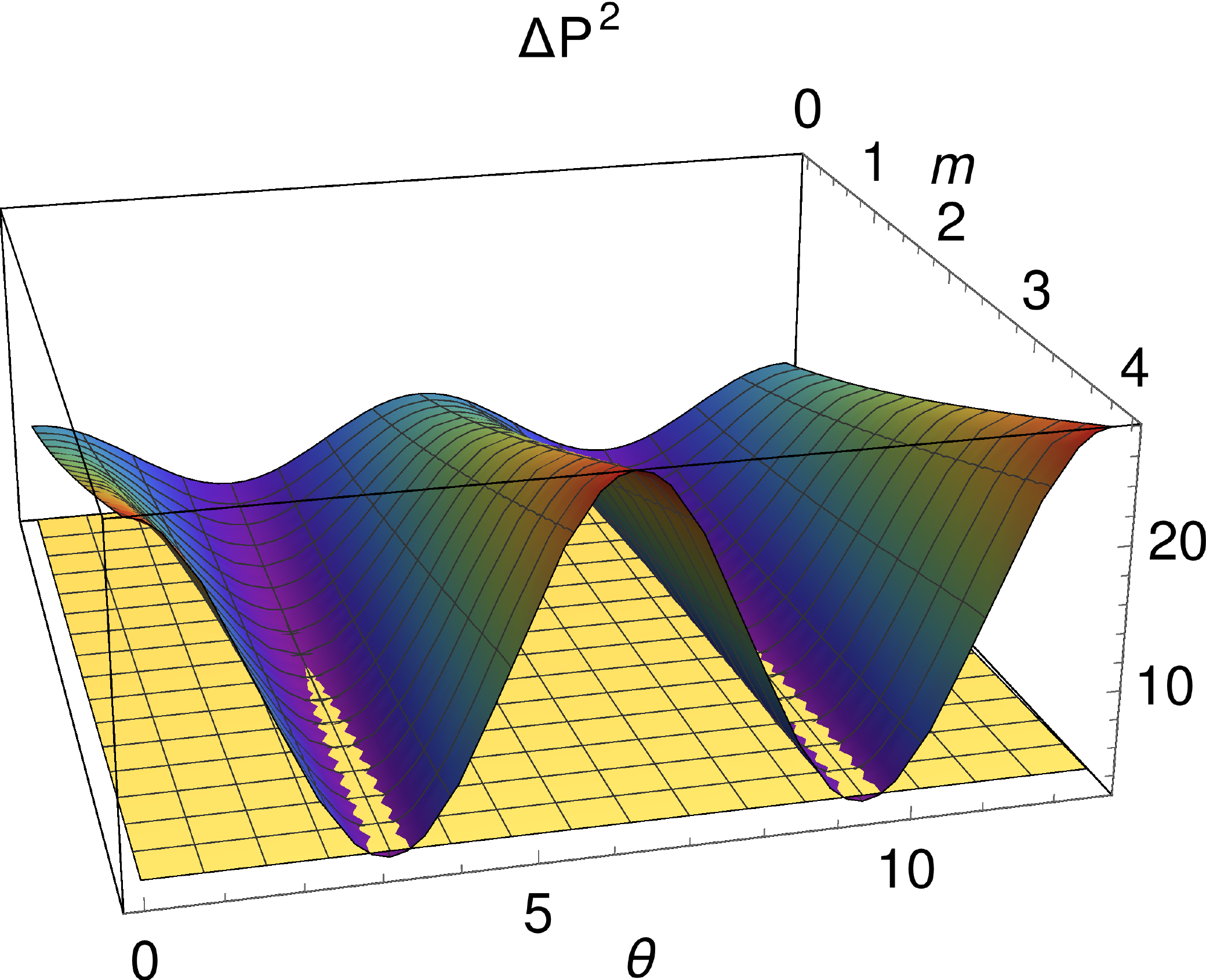}
    \label{fig4e}} 
\subfigure[]{
  \includegraphics[width=5.7cm]{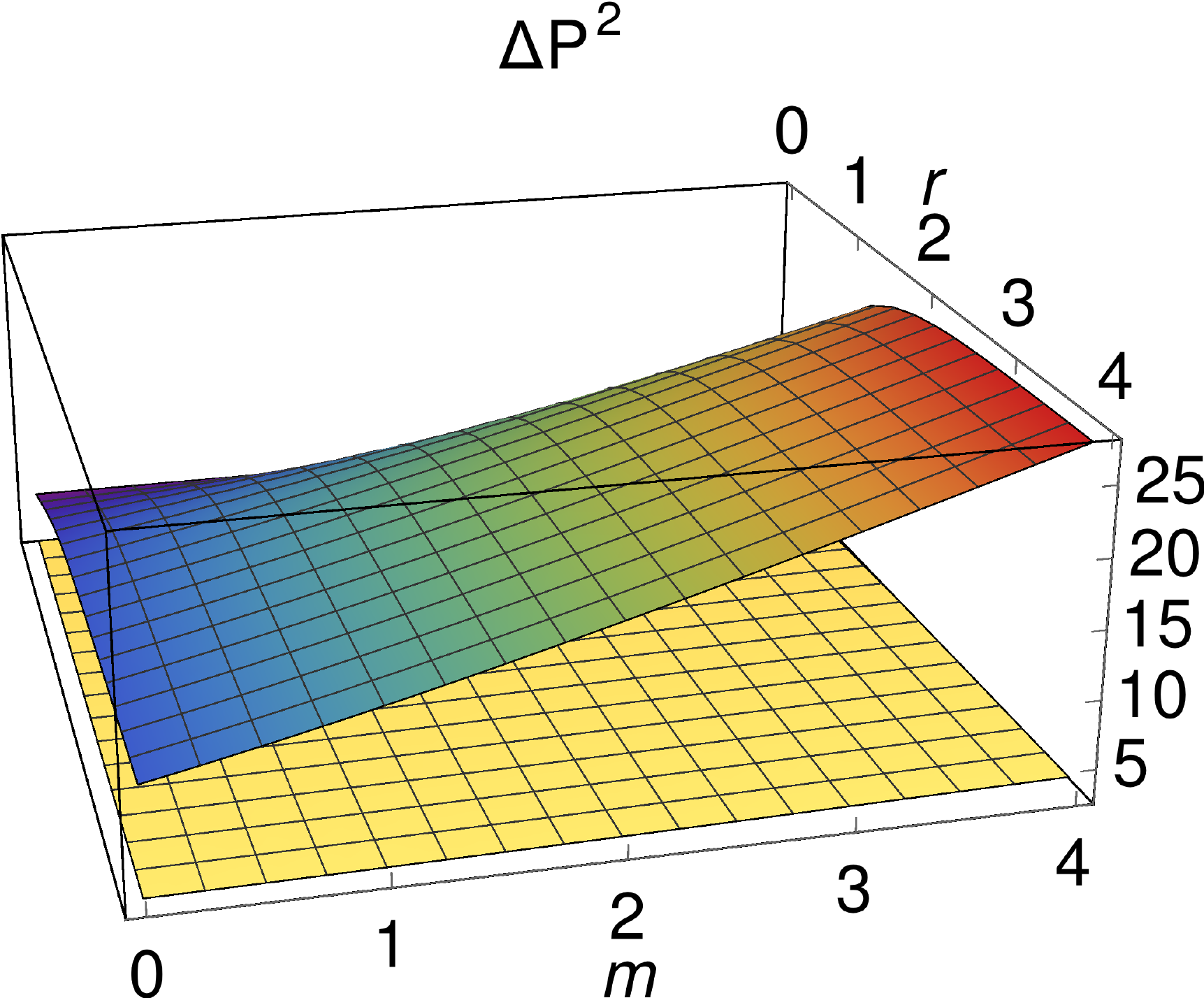}
  \label{fig4f}}
 \caption{(Color online) Square of the uncertainties of $X$ and $P$ quadratures for odd generalized PSSVS for $m=3$ in \subref{fig4a} $\&$ \subref{fig4d}, $r=5$ in \subref{fig4b} $\&$ \subref{fig4e} and $\theta = 0$ in \subref{fig4c} $\&$ \subref{fig4f}. In all the plots $\lambda$ and $\kappa$ are chosen to be $1.5$. The yellow surfaces in all the figures show the variation of the RHS of the generalized Robertson uncertainty relation.}
  \label{QS_odd}
\end{figure*}
\section{The model}\label{sec3}
The formalism that we have proposed here is entirely general and, thus, it can be applied to any quantum mechanical, nonlinear or even to any deformed quantum mechanical scenarios. The only task is to extract the value of the function $f(n)$ from the given model. In this article, we shall utilize the standard solutions of the trigonometric P\"oschl-Teller potential and show how one can construct the PSSVS for such a systems and explore their nonclassical properties. The explicit form of the potential is \cite{Antoine_Gazeau_Monceau_Klauder_Penson}
\begin{equation}\label{Potential}
V(x)=\frac{\hbar^2}{8ma^2}\left[\frac{\lambda(\lambda-1)}{\cos^2\frac{x}{2a}}+\frac{\kappa(\kappa-1)}{\sin^2\frac{x}{2a}}-(\lambda+\kappa)^2\right],
\end{equation}
and the corresponding eigenvalues and eigenfunctions are given by
\begin{eqnarray}
&& \psi_n(x)=\mathcal{N}(\kappa,\lambda)\left(\cos\frac{x}{2a}\right)^\lambda\left(\sin\frac{x}{2a}\right)^\kappa{}_\\
&& \qquad\qquad \times {}_2F_1\left(-n,n+\lambda+\kappa;\kappa+\frac{1}{2};\sin^2\frac{x}{2a}\right), \notag \\
&& E_n=\frac{\hbar^2}{2ma^2}n(n+\lambda+\kappa). \label{eigenvalues}
\end{eqnarray}
The above solutions are readily available in the literature, see; for instance, \cite{Antoine_Gazeau_Monceau_Klauder_Penson}. One of the reasons behind choosing such a model for our analysis is that it is an interesting variant of potential well describing many physical systems, particularly, many body systems in atomic and molecular physics. P\"oschl-Teller potential is a more interesting model compared to the usual infinite potential well, since it can provide deeper insights to some important formalisms \cite{Reed,Antoine_Gazeau_Monceau_Klauder_Penson}, such as; defining the domain of the self-adjoint operators that are used in constructing the potential, non-uniqueness of self-adjoint extensions, semi-classical nature as well as the classical limit, exotic roles played by the boundary conditions, etc. Moreover, the potential \myref{Potential} comes out in a general form, and depending on the values of the parameters $\lambda$ and $\kappa$, one can modify it in different ways such that it becomes useful in various contexts. For instance, if one restricts $1/2\leq\lambda<1$, one obtains an interesting potantial being familiar as the Scarf potential, which belongs to the category of inverted potential well. For further information in this regard, one may refer; for instance \cite{Antoine_Gazeau_Monceau_Klauder_Penson}. 

In our analysis we have chosen $\lambda=\kappa=1.5$ so that the potential \myref{Potential} turns out to be a symmetric P\"oschl-Teller potential. As per our formalism, we can assume that the Hamiltonian can be factorized in terms of the generalized ladder operators \myref{GenLad} as $\hat{H}=\hat{A}^\dagger \hat{A}$ to reproduce the spectrum \myref{eigenvalues}. Thus, the number operator can be considered as $\hat{N}'=\hat{A}^\dagger \hat{A}=\hat{n}f^2(\hat{n})$, with
\begin{equation}\label{fn}
f(n)=(n+\lambda+\kappa)^{1/2},
\end{equation}
which when replaced in \myref{EPSSVS} and \myref{OPSSVS} one obtains the even and odd PSSVS, respectively, for the corresponding model. Note that, we have chosen the values of $\hbar,m$ and $a$ in such a way that the factor $\hbar^2/2ma^2$ in \myref{eigenvalues} becomes unity. It is possible to realize the number operator in an alternative way also, for instance, we can define it in the following way 
\begin{equation}\label{NumOp}
\hat{N}=\left[\hat{A}^\dagger \hat{A}+\frac{1}{4}(\lambda+\kappa)^2\right]^{1/2}-\frac{1}{2}(\lambda+\kappa).
\end{equation}
There are several advantages against such a choice, firstly, $\hat{N}$ in \myref{NumOp} becomes equivalent to the usual photon number operator $\hat{n}=\hat{a}^\dagger \hat{a}$ in the sense that $\hat{N}|n\rangle=\hat{n}|n\rangle=n|n\rangle$. That means although the information of the generalized model is hidden inside $\hat{N}$ in terms of $f(n)$, it still produces a physical photon, whereas $\hat{N}'=\hat{A}^\dagger A$ does not. Owing to such an interesting physical outcome, we are more interested in choosing the number operator given by \myref{NumOp} in the rest of our analysis. Interestingly, we observe that the Hamiltonian can be factorized in terms of the new number operator $\hat{N}$ also as $\hat{H}=\hat{N}(\hat{N}+\lambda+\kappa)=\hat{A}^\dagger \hat{A}=\hat{N}f^2(\hat{N})$, with
\begin{equation}\label{fN}
f(N)=(N+\lambda+\kappa)^{1/2},
\end{equation}
so that one can identify $\hat{N}$ to be equivalent to $\hat{n}$. Therefore, while choosing the function $f(n)$ it becomes immaterial whether one takes it from \myref{fn} or \myref{fN}, physically they are same.
\section{Signature of nonclassicality}\label{sec4}
Any states which are less classical than the coherent states are familiar as nonclassical states. In other words, nonclassicality is a measure of quantumness of a state with reference to the Glauber coherent state. In our analysis, our purpose is to explore the nonclassical properties of generalized even and odd PSSVS and, we shall quantify the nonclassicality with respect to the Glauber coherent states. Generalized coherent states are already known to be nonclassical \cite{Dey_Fring_squeezed,Dey_Fring_Gouba_Castro,Dey_Fring_Hussin} while measuring it with respect to the Glauber coherent states, thus, we do not measure the nonclassicality of generalized PSSVS with respect to the generalized coherent states.

Among various existing methods of measuring nonclassicality, here we choose only three, namely quadrature squeezing, photon number squeezing and Wigner quasi-probability distribution function, which are based on three independent frameworks. A quantum optical state is defined to be nonclassical if its nonclassicality is identified from at least one of the standard methods. It is not always guaranteed that if a state is nonclassical based on a given framework, it will be nonclassical with respect to the other methods. For instance, in \cite{Dey}, it was shown that the quadratures of $q$-deformed coherent states are not squeezed, but the photon distribution is sub-Poissonian. Thus, the corresponding state was argued to be weakly nonclassical compared to the Sch\"odinger cat state, which was also explored in the same paper. Therefore, it is legitimate to claim that if the nonclassicality of a state is detected from several independent methods, it is more convincing that the state belongs to the category of stronger nonclassical states. We analyze the nonclassical properties of generalized PSSVS arising from the three methods as discussed in the following subsections. Since our state originates from the generalized quantum optical framework, obviously we require some modifications of the standard protocols of each of the procedures, which we discuss in detail in the respective subsections.
\subsection{Squeezing in quadratures}\label{sec31}
First, we define the quadrature operators for the generalized system as $\hat{X}=(\hat{A}+\hat{A}^\dagger)/\sqrt{2}$ and $\hat{P}=(\hat{A}-\hat{A}^\dagger)/\sqrt{2}i$ so that they obey the generalized Robertson uncertainty relation
\begin{equation}\label{GUR}
\Delta \hat{X}\Delta \hat{P}\geq\frac{1}{2}\Big\vert\langle\cdot|[\hat{X},\hat{P}]|\cdot\rangle\Big\vert.
\end{equation}
We are bound to replace the standard definition of the quadrature operators by the new definition as mentioned above, since any observable out of our generalized system has to be composed of the generalized ladder operators by construction. This, in turn, demands a modification of the definition of the quadrature squeezing itself, since we have to deal with the factor $f(n)$ that are associated with the generalized ladder operators. The replacement of Heisenberg uncertainty relation by the Roberson uncertainty relation would accomplish the job, as the RHS of generalized uncertainty relation \myref{GUR} consists of the commutator $[\hat{X},\hat{P}]$ within which the signature of the function $f(n)$ is underlain. While computing \myref{GUR} using the vacuum state, the square root of the RHS of \myref{GUR} and each of the variances turns out to be equal, viz.\,$\Delta \hat{X}=\Delta \hat{P}=\sqrt{|\langle 0|[\hat{X},\hat{P}]|0\rangle|/2}=\sqrt{f(1)/2}$, and the same result can be obtained using the deformed coherent states also \cite{Dey}. However, this may be true only for certain coherent states and it is not guaranteed to hold for any arbitrary generalized state. This is because the commutator $[\hat{X},\hat{P}]$ is not proportional to the identity operator in general, thus, the result of the RHS of \myref{GUR} is dependent on the state that is used to compute the expectation values. Therefore, it is not guaranteed that the variances corresponding to the generalized PSSVS are necessarily below to those of the vacuum state, as it happens in the case of harmonic oscillator. In our case, the quadrature squeezing does not correspond to the case when the variance of any of the quadratures becomes lower than the square root of the RHS of \myref{GUR} for the vacuum state, but, it corresponds to that of the particular state that is being studied, which is $|\zeta,f,m\rangle_{\text{e/o}}$ in our case.

In order to compute the quadrature squeezing analytically, we first need to compute the expectation values of ladder operators $\hat{A},\hat{A}^\dagger$ and their squares using the states $|\zeta,f,m\rangle_{\text{e/o}}$. The expectation values of $\hat{A}$ and $\hat{A}^\dagger$ for both even and odd PSSVS can clearly be understood to be vanished, because, $\hat{A}$ operating on the even/odd state will make it odd/even, which when is overlapped with the even/odd state will surely lead to zero. The same reasoning can be applied to $\hat{A}^\dagger$. Thus,
\begin{figure*}
  \subfigure[]{
\includegraphics[width=7.5cm,height=4.8cm]{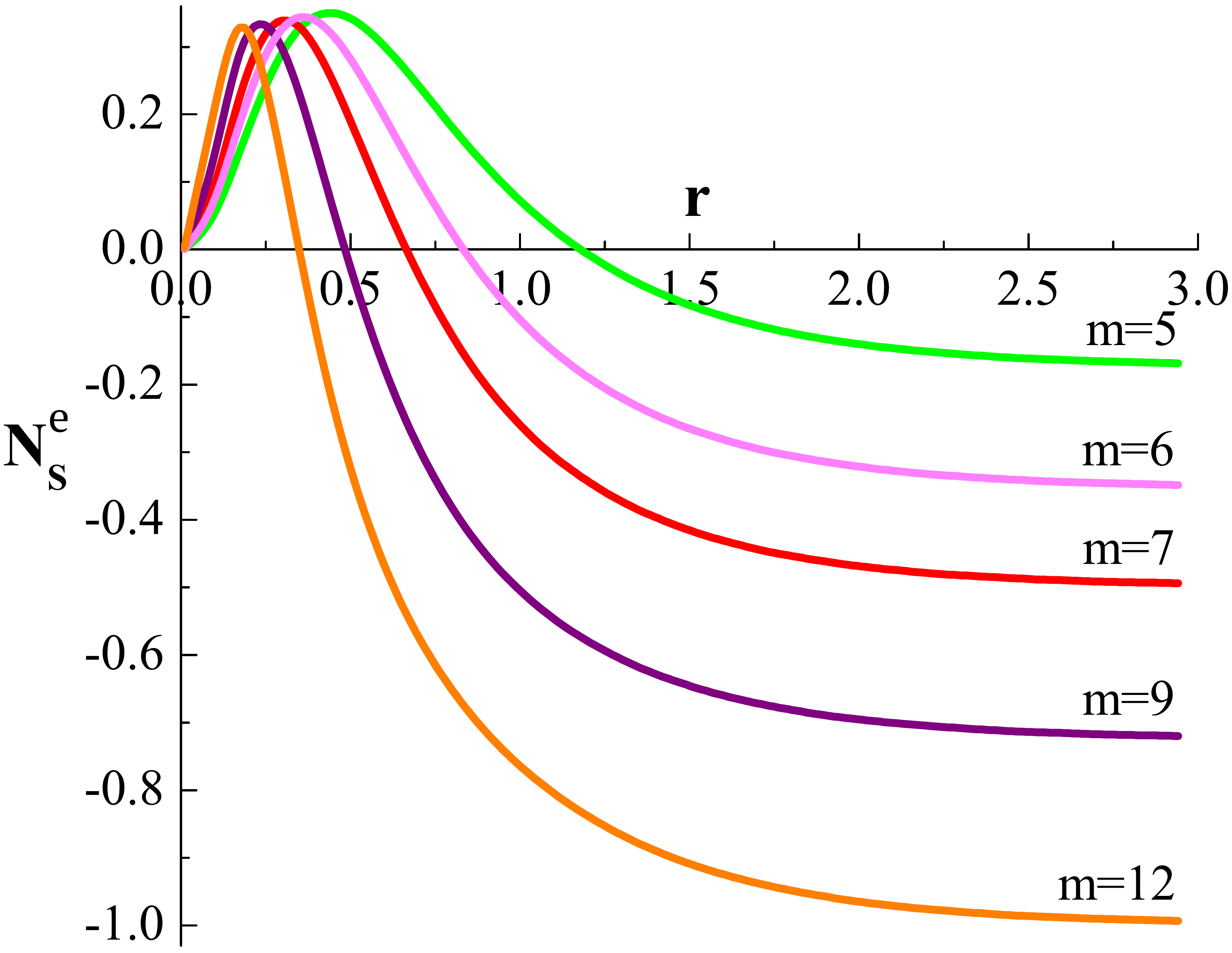}
\label{fig1a}}
  \subfigure[]{
    \includegraphics[width=7.5cm,height=4.8cm]{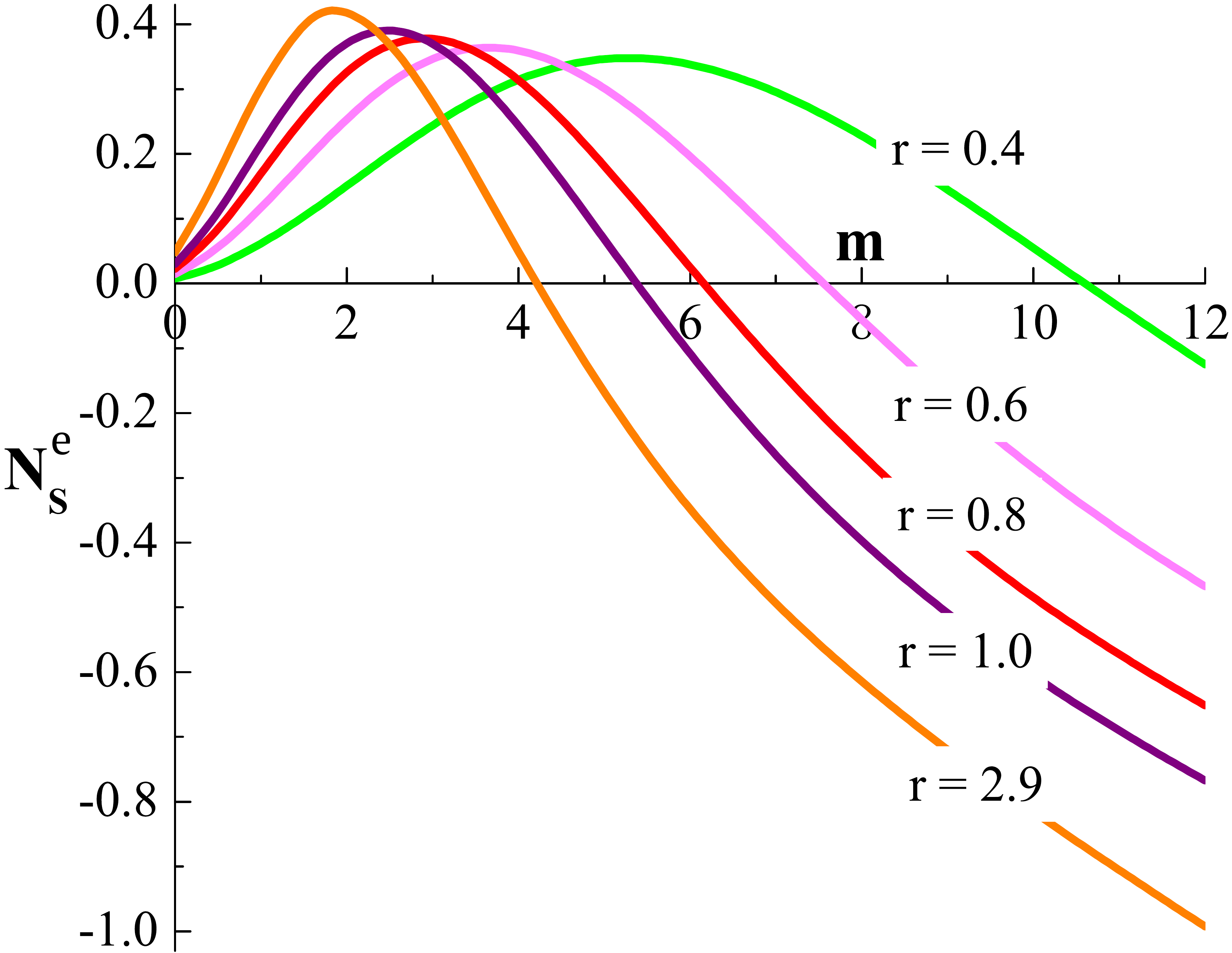}
    \label{fig1b}} 
\subfigure[]{
  \includegraphics[width=7.5cm,height=4.8cm]{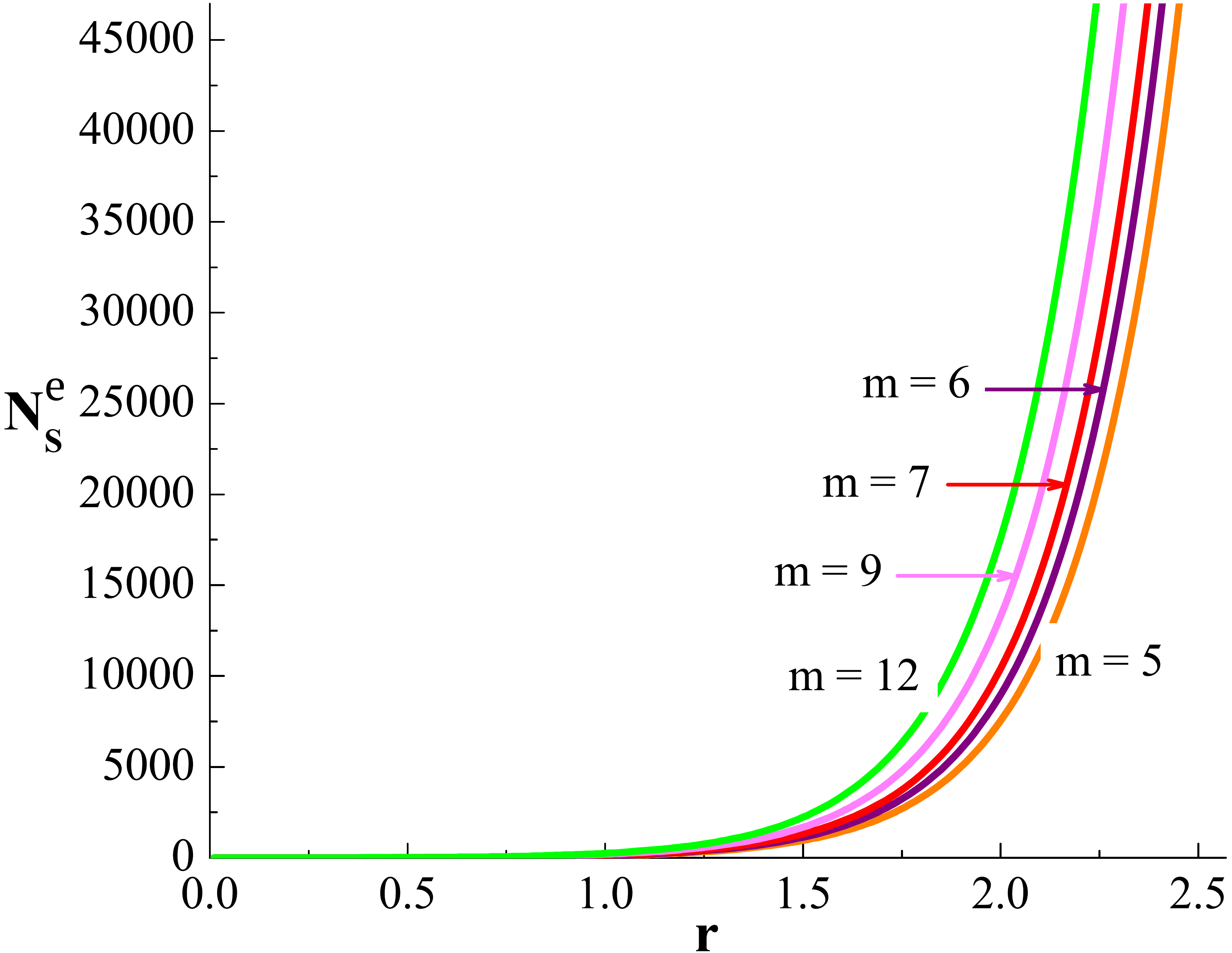}
  \label{fig1c}}
\subfigure[]{
  \includegraphics[width=7.5cm,height=4.8cm]{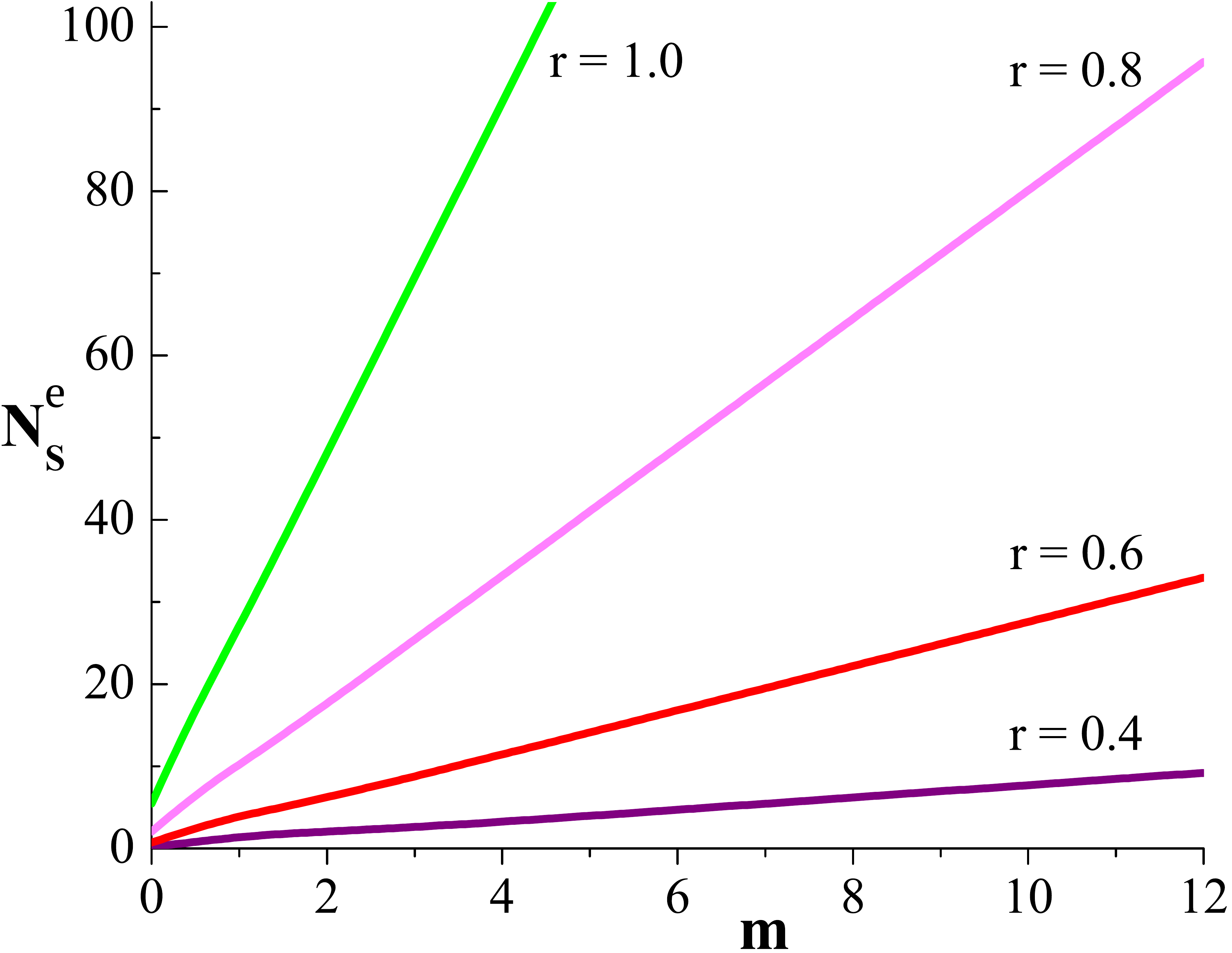}
  \label{fig1d}}
 \caption{(Color online) Number squeezing $N_\text{s}^{\text{e}}$ for even PSSVS for generalized cases $f(n)=\sqrt{n+\kappa+\lambda}$ in \subref{fig1a} and \subref{fig1b} for $\lambda=\kappa=1.5$, and for harmonic oscillator cases $f(n)=1$ in \subref{fig1c} and \subref{fig1d}.}
  \label{NS_even}
\end{figure*}
\begin{eqnarray}
&& {}_{\text{e/o}}\langle\zeta,f,m|\hat{A}|\zeta,f,m\rangle_{\text{e/o}}=0, \\
&& {}_{\text{e/o}}\langle\zeta,f,m|\hat{A}^\dagger|\zeta,f,m\rangle_{\text{e/o}}=0.
\end{eqnarray}
The expectation values of the squares of $\hat{A},\hat{A}^\dagger$ are computed as follows
\begin{alignat}{1}
& {}_{\text{e}}\langle \hat{A}^2\rangle_{\text{e}}=\frac{-e^{i\theta}}{\mathcal{N}^{\text{e}^2}_{\zeta,f,m}}\displaystyle\sum_{n=0}^\infty\frac{(\tanh r/2)^{2k+1}(2k)!(2k+2)!}{k!(k+1)!(2n)![f(2n)!]^2},\\
& {}_{\text{o}}\langle \hat{A}^2\rangle_{\text{o}}=\frac{-e^{i\theta}}{\mathcal{N}^{\text{o}^2}_{\zeta,f,m}}\displaystyle\sum_{n=0}^\infty\frac{(\tanh r/2)^{2k+3}(2k+2)!(2k+4)!}{(k+1)!(k+2)!(2n+1)![f(2n+1)!]^2},
\end{alignat}
with ${}_{\text{e}}\langle \hat{A}^{\dagger^2}\rangle_{\text{e}}={}_{\text{e}}\langle \hat{A}^2\rangle_{\text{e}}^\ast$ and ${}_{\text{o}}\langle \hat{A}^{\dagger^2}\rangle_{\text{o}}={}_{\text{o}}\langle \hat{A}^2\rangle_{\text{o}}^\ast$. Finally, we compute
\begin{alignat}{1}
& {}_{\text{e}}\langle \hat{A}\hat{A}^\dagger\rangle_{\text{e}}=\frac{1}{\mathcal{N}^{\text{e}^2}_{\zeta,f,m}}\displaystyle\sum_{n=0}^\infty\frac{2^{-2k}[(2k)!]^2(2n+1)[f(2n+1)]^2}{(\tanh r)^{-2k}[k!]^2(2n)![f(2n)!]^2}, \\
& {}_{\text{e}}\langle \hat{A}^{\dagger}\hat{A}\rangle_{\text{e}}=\frac{1}{\mathcal{N}^{\text{e}^2}_{\zeta,f,m}}\displaystyle\sum_{n=0}^\infty\frac{(\tanh r/2)^{2k+2}[(2k+2)!]^2}{[(k+1)!]^2(2n+1)![f(2n+1)!]^2}, \label{AAdag1} \\
& {}_{\text{o}}\langle \hat{A}\hat{A}^\dagger\rangle_{\text{o}}=\frac{1}{\mathcal{N}^{\text{o}^2}_{\zeta,f,m}}\displaystyle\sum_{n=0}^\infty\left[\frac{(\tanh r)^{2k+2}[(2k+2)!]^2}{2^{2k+2}[(k+1)!]^2(2n+1)!}\right.\\
&\qquad\qquad\qquad\qquad\qquad\quad  \times\left.\frac{(2n+2)[f(2n+2)]^2}{[f(2n+1)!]^2} \right],\notag\\
& {}_{\text{o}}\langle \hat{A}^{\dagger}\hat{A}\rangle_{\text{o}}=\frac{1}{\mathcal{N}^{\text{o}^2}_{\zeta,f,m}}\displaystyle\sum_{n=0}^\infty\frac{(\tanh r/2)^{2k+2}[(2k+2)!]^2}{[(k+1)!]^2(2n)![f(2n)!]^2}, \label{AAdag2}
\end{alignat}
so that the variance of the quadratures $\hat{X}$ and $\hat{P}$ can be calculated as $(\Delta \hat{X}_{\text{e/o}})^2={}_{\text{e/o}}\langle \hat{X}^2\rangle_{\text{e/o}}=({}_{\text{e/o}}\langle \hat{A}\hat{A}^\dagger\rangle_{\text{e/o}}+{}_{\text{e/o}}\langle \hat{A}^\dagger \hat{A}\rangle_{\text{e/o}}+{}_{\text{e/o}}\langle \hat{A}^2\rangle_{\text{e/o}}+{}_{\text{e/o}}\langle \hat{A}^{\dagger^2}\rangle_{\text{e/o}})/2$ and $(\Delta \hat{P}_{\text{e/o}})^2={}_{\text{e/o}}\langle \hat{P}^2\rangle_{\text{e/o}}=({}_{\text{e/o}}\langle \hat{A}\hat{A}^\dagger\rangle_{\text{e/o}}+{}_{\text{e/o}}\langle \hat{A}^\dagger \hat{A}\rangle_{\text{e/o}}-{}_{\text{e/o}}\langle \hat{A}^2\rangle_{\text{e/o}}-{}_{\text{e/o}}\langle \hat{A}^{\dagger^2}\rangle_{\text{e/o}})/2$, respectively. The numerical studies of the variances for $\hat{X}$ and $\hat{P}$ quadratures with respect to different parameters are depicted in Fig.\,\ref{QS_even} for even states and in Fig.\,\ref{QS_odd} for the odd states. The yellow surfaces in each of the plots represent the variation of the RHS of the generalized Robertson uncertainty relation. Thus, any portion of the variances of the quadratures which falls below the yellow surface clearly identifies the squeezing of quadrature. We observe that the quadratures are squeezed in all the cases (at least partially in certain regime) except in Fig.\,\ref{fig3f} and \ref{fig4f}, where we have fixed the value of $\theta=0$. However, this is purely because of the choice of $\theta$. If we would have chosen the value of $\theta$ at around $4$ (in the appropriate unit) to re-plot the Fig.\,\ref{QS_even}\subref{fig3f}, we would have obtained the squeezing in $\hat{P}$ quadrature also. This can be ensured by a careful observation of Fig.\,\ref{QS_even}\subref{fig3e}. A similar type of argument is also true for the Fig.\,\ref{QS_odd}\subref{fig4f}. Nevertheless, we notice that the variation of the quadrature is periodic in $\theta$ and one must choose the value of the parameter $\theta$ appropriately to obtain the squeezing in both of the quadratures (of course, not at the same time, but alternatively).  
\begin{figure*}
  \subfigure[]{
\includegraphics[width=7.5cm,height=4.8cm]{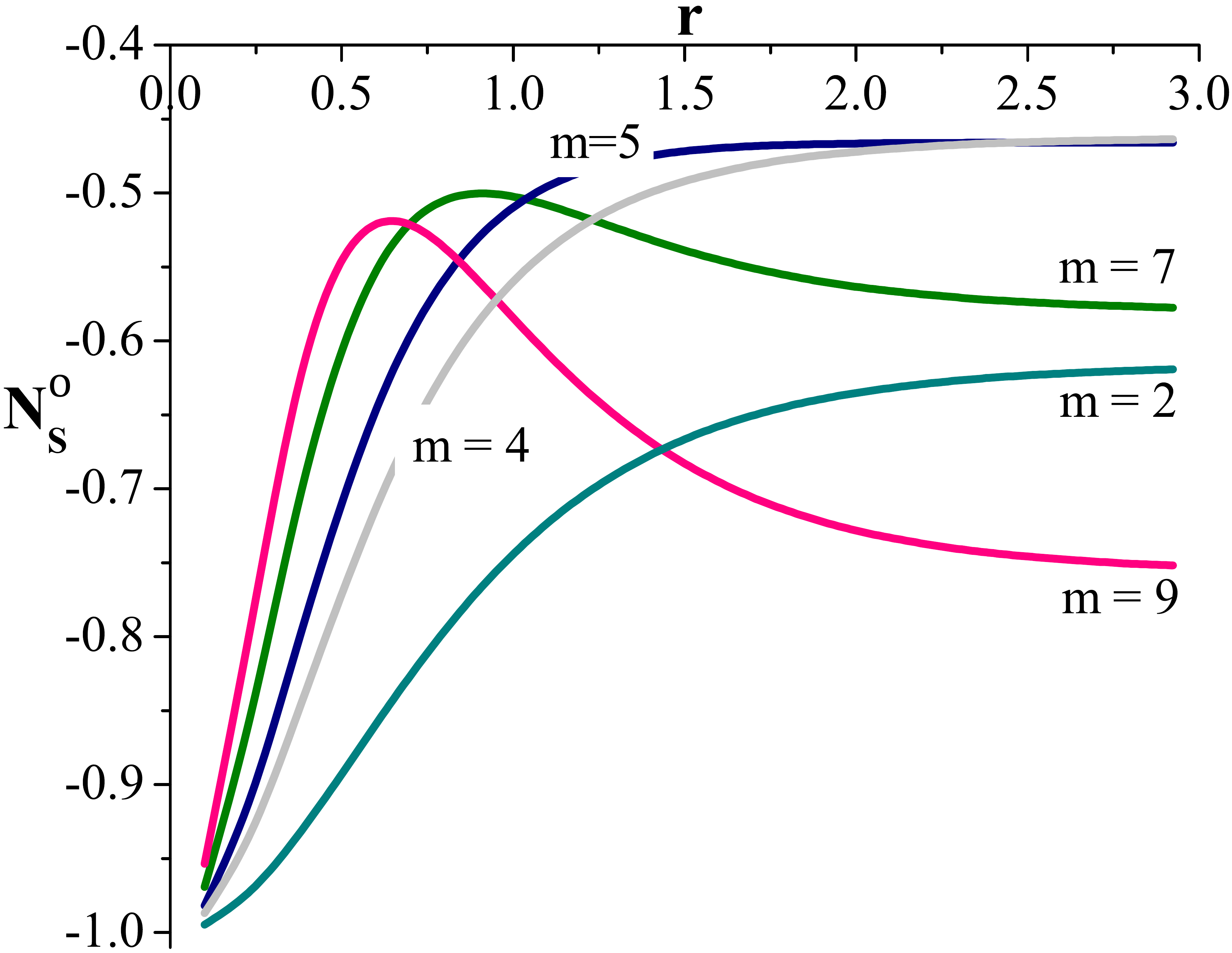}
\label{fig2a}}
  \subfigure[]{
    \includegraphics[width=7.5cm,height=4.8cm]{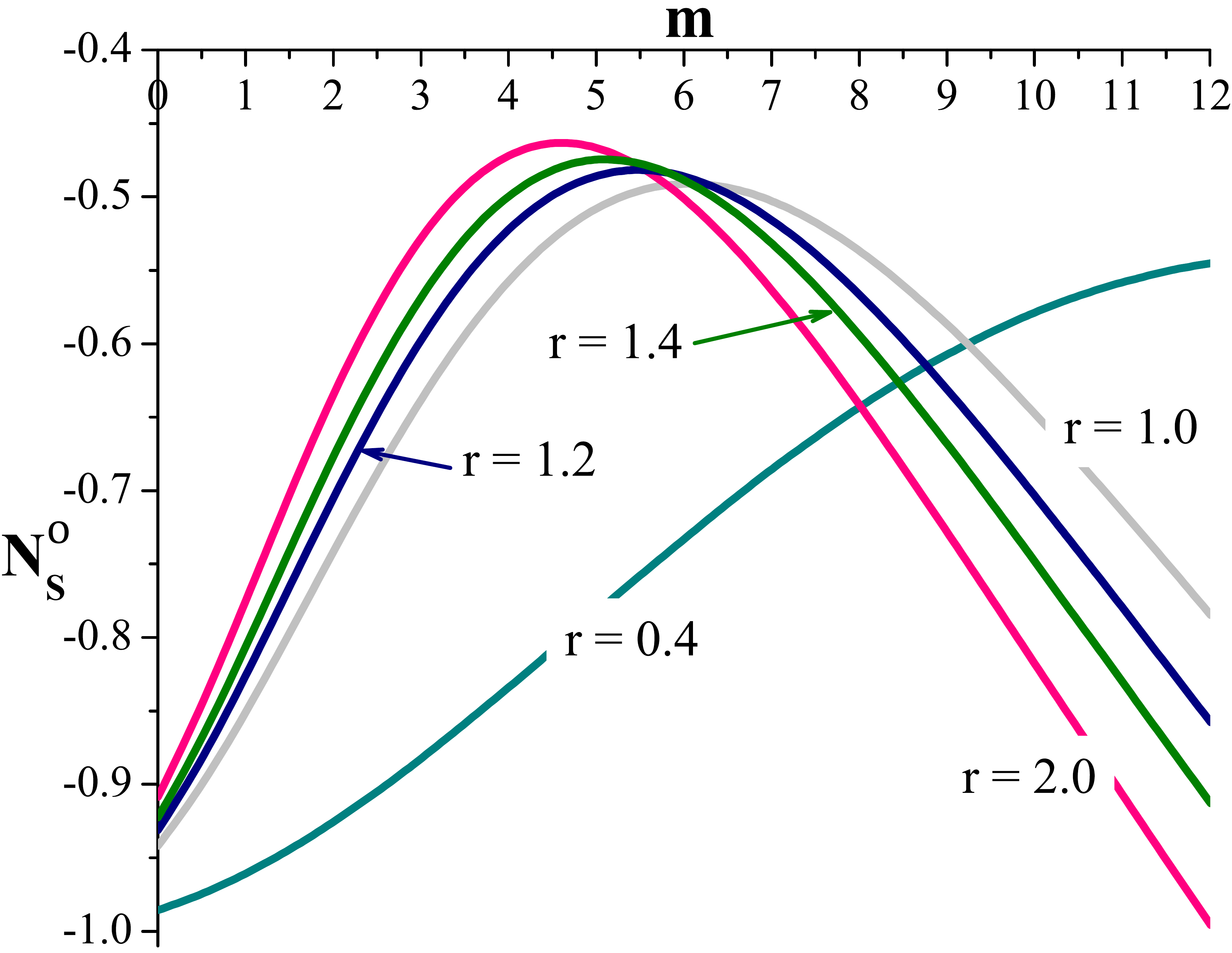}
    \label{fig2b}} 
\subfigure[]{
  \includegraphics[width=7.5cm,height=4.8cm]{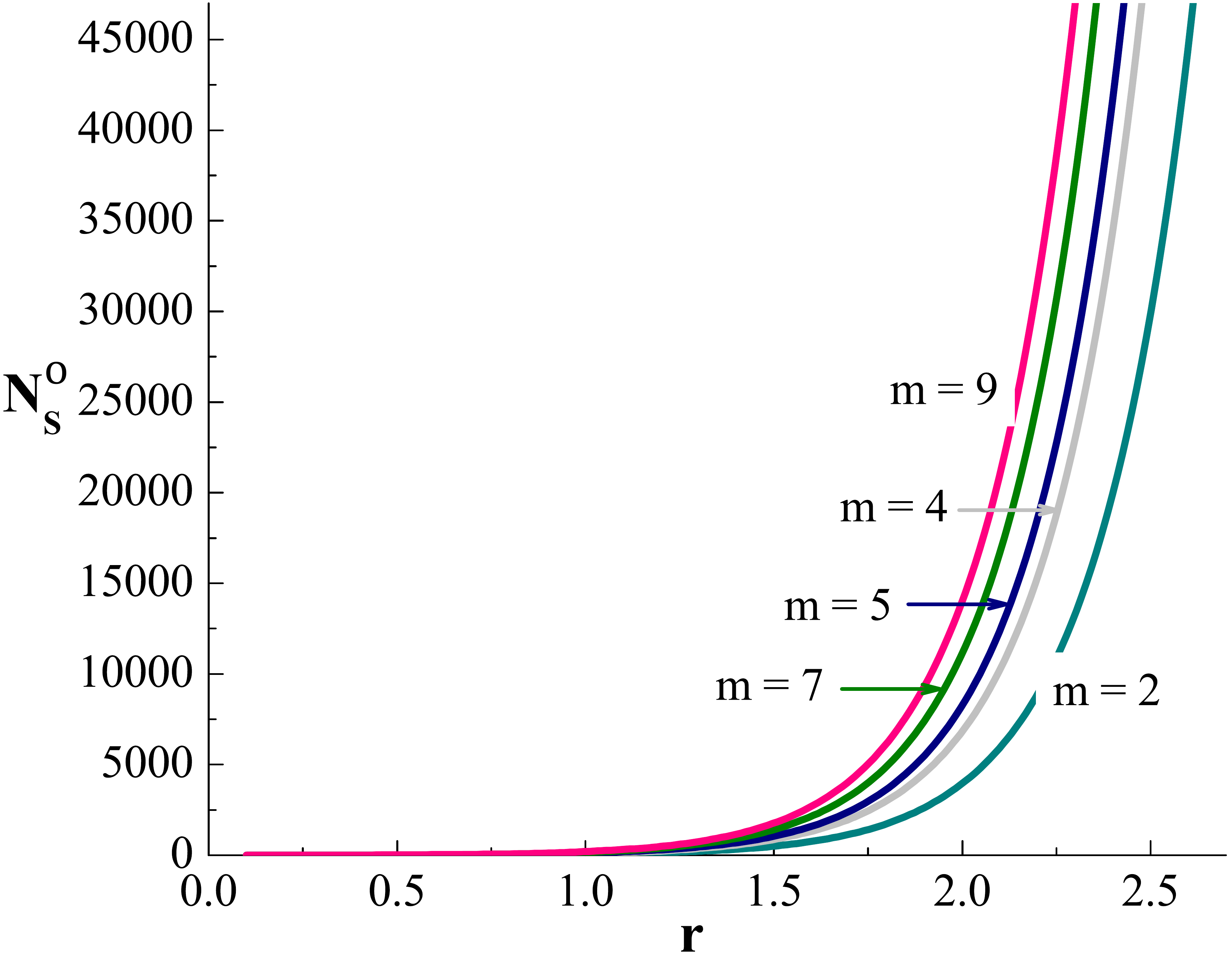}
  \label{fig2c}}
\subfigure[]{
  \includegraphics[width=7.5cm,height=4.8cm]{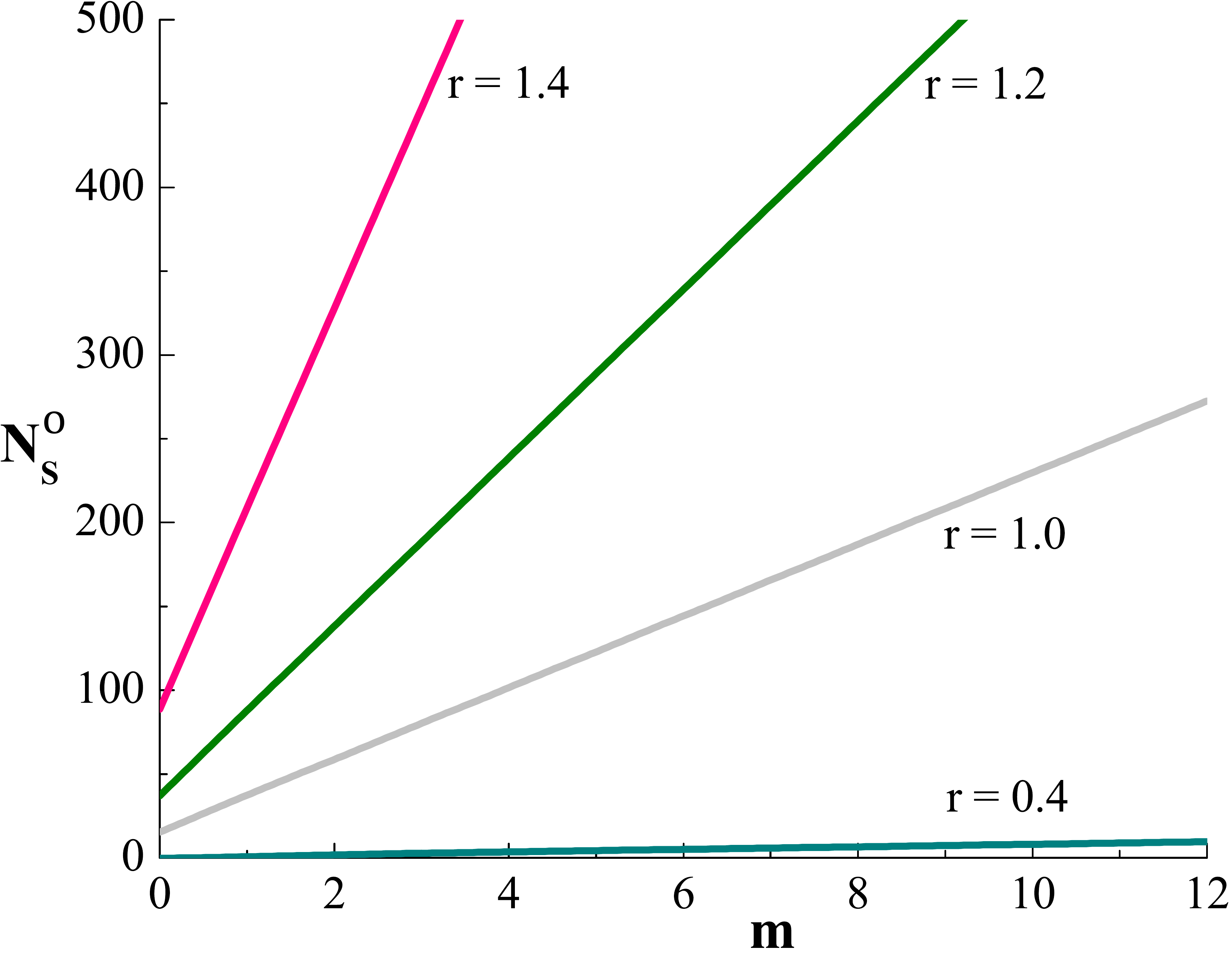}
  \label{fig2d}}
 \caption{(Color online) Number squeezing $N_\text{s}^{\text{o}}$ for odd PSSVS for generalized cases $f(n)=\sqrt{n+\kappa+\lambda}$ in \subref{fig2a} and \subref{fig2b} for $\lambda=\kappa=1.5$, and for harmonic oscillator cases $f(n)=1$ in \subref{fig2c} and \subref{fig2d}.}
  \label{NS_odd}
\end{figure*}
\subsection{Photon statistics}\label{sec32}
The study of photon statistics in the generalized case also requires a modification over the standard method. Usually the photon number of a quantum optical state is said to be squeezed if the distribution function corresponding to the state is sub-Poissonian and the signature of nonclassicality/squeezing is easily captured by utilizing the Mandel parameter $Q=\langle (\Delta \hat{n})^2\rangle/\langle \hat{n}\rangle-1$ \cite{Mandel}. Notice that when the distribution is Poissonian, i.e. $\langle (\Delta \hat{n})^2\rangle=\langle \hat{n}\rangle$ (which is also the characteristics of Glauber coherent states), the Mandel parameter turns out to be zero. A negative Mandel parameter indicates a sub-Poissonian statistics of the photon distribution. Thus, a simple computation of the Mandel parameter can ensure the nature of the photon distribution of any quantum optical state that originates from the bosonic ladder operators. 

A straightforward generalization of the Mandel parameter is carried out by replacing the bosonic number operator $\hat{n}$ with the generalized number operator $\hat{N}$ as $Q_{\text{g}}=\langle (\Delta \hat{N})^2\rangle/\langle \hat{N}\rangle-1$. Alternatively, one can check whether the condition  $\langle (\Delta \hat{N})^2\rangle <\langle \hat{N}\rangle$ is valid or not. In other words, we can compute $N_{\text{s}}=\langle (\Delta \hat{N})^2\rangle -\langle \hat{N}\rangle$, and if it turns out to be negative, we can affirm the nonclassicality of the state. Note that, at this point, people often get confused whether the nonclassicality, within this formalism, is computed with respect to the generalized coherent states or with respect to the Glauber coherent states. It is already known that the generalized coherent states possess sub-Poissonian photon statistics \cite{Dey,Dey_Fring_Hussin,Zelaya} with the corresponding Mandel parameter being $Q_g=\langle[\hat{A},\hat{A}^\dagger]\rangle-1$. It is also true that $Q_g$ is non-zero if we compute it in the Glauber coherent state basis. Thus, it will be inappropriate to compare the nonclassicality of generalized PSSVS with respect to any of the two types of coherent states. It is, rather, more evident that the nonclassicality here is computed with respect to a state whose photon distribution is Poissonian (Glauber coherent state in the standard harmonic oscillator basis being one of the examples). With this note,  we
\begin{figure*}
  \subfigure[]{
\includegraphics[width=7.5cm,height=5cm]{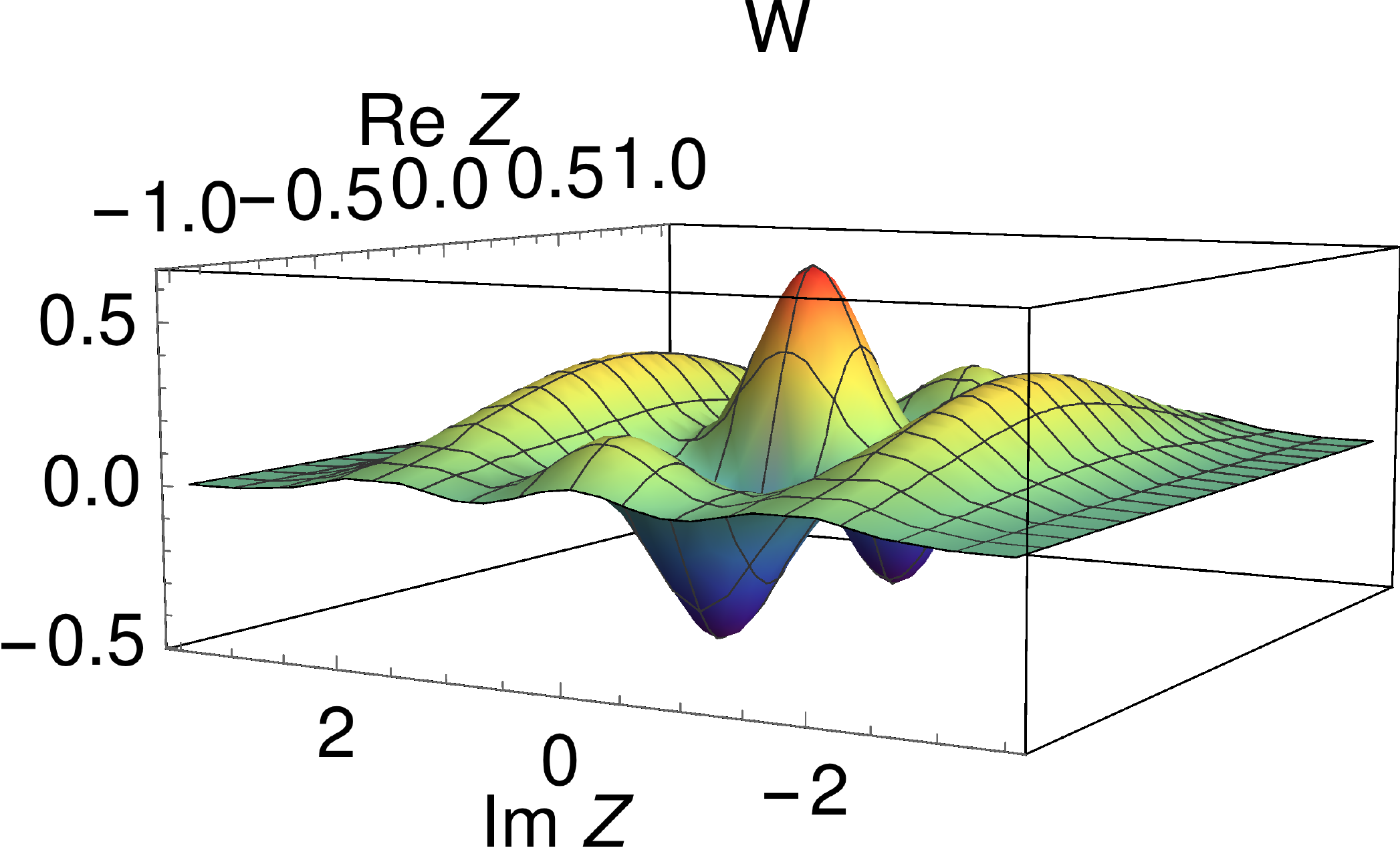}
\label{fig5a}}
  \subfigure[]{
    \includegraphics[width=7.5cm,height=5cm]{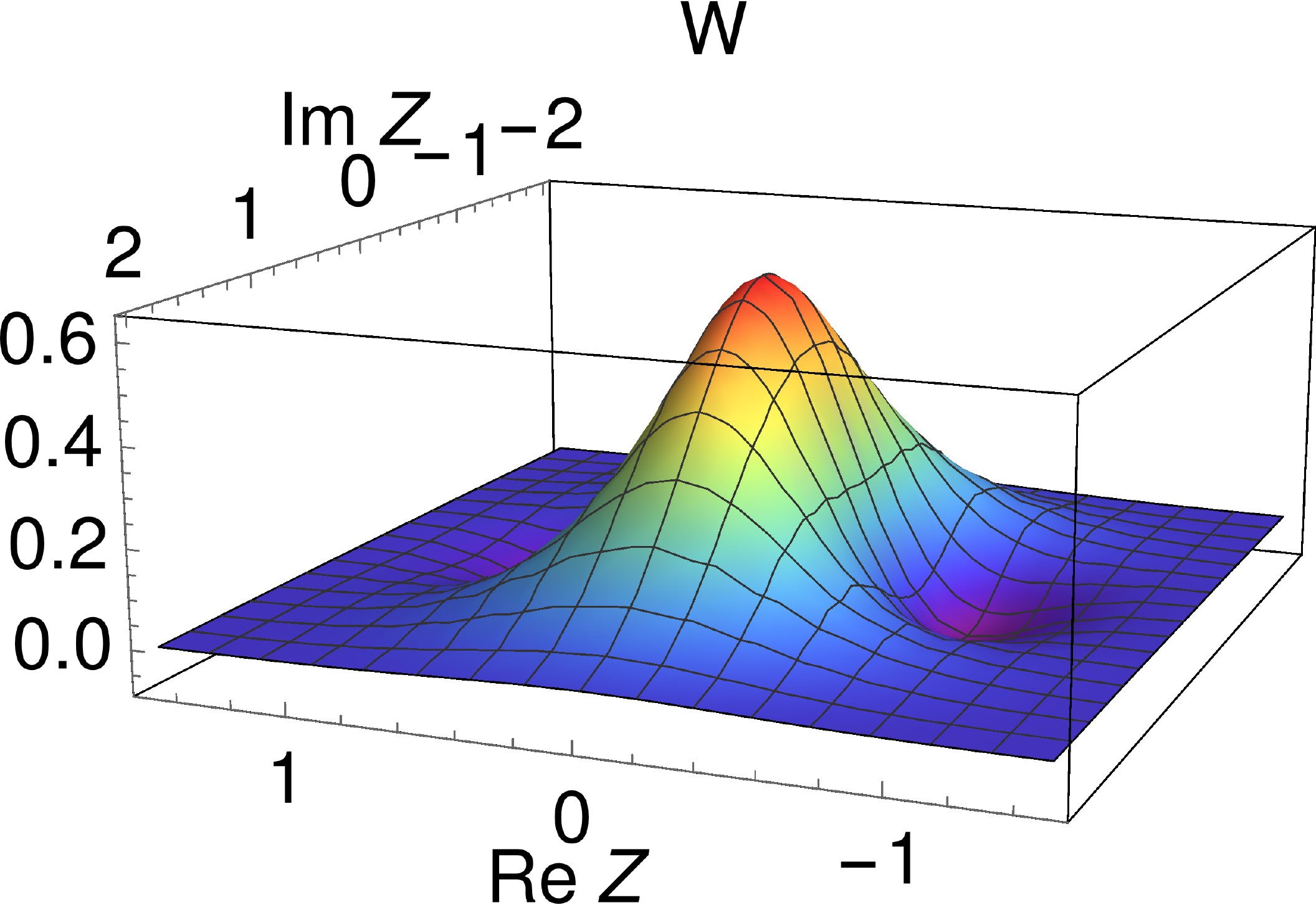}
    \label{fig5b}} 
\subfigure[]{
  \includegraphics[width=7.5cm,height=5cm]{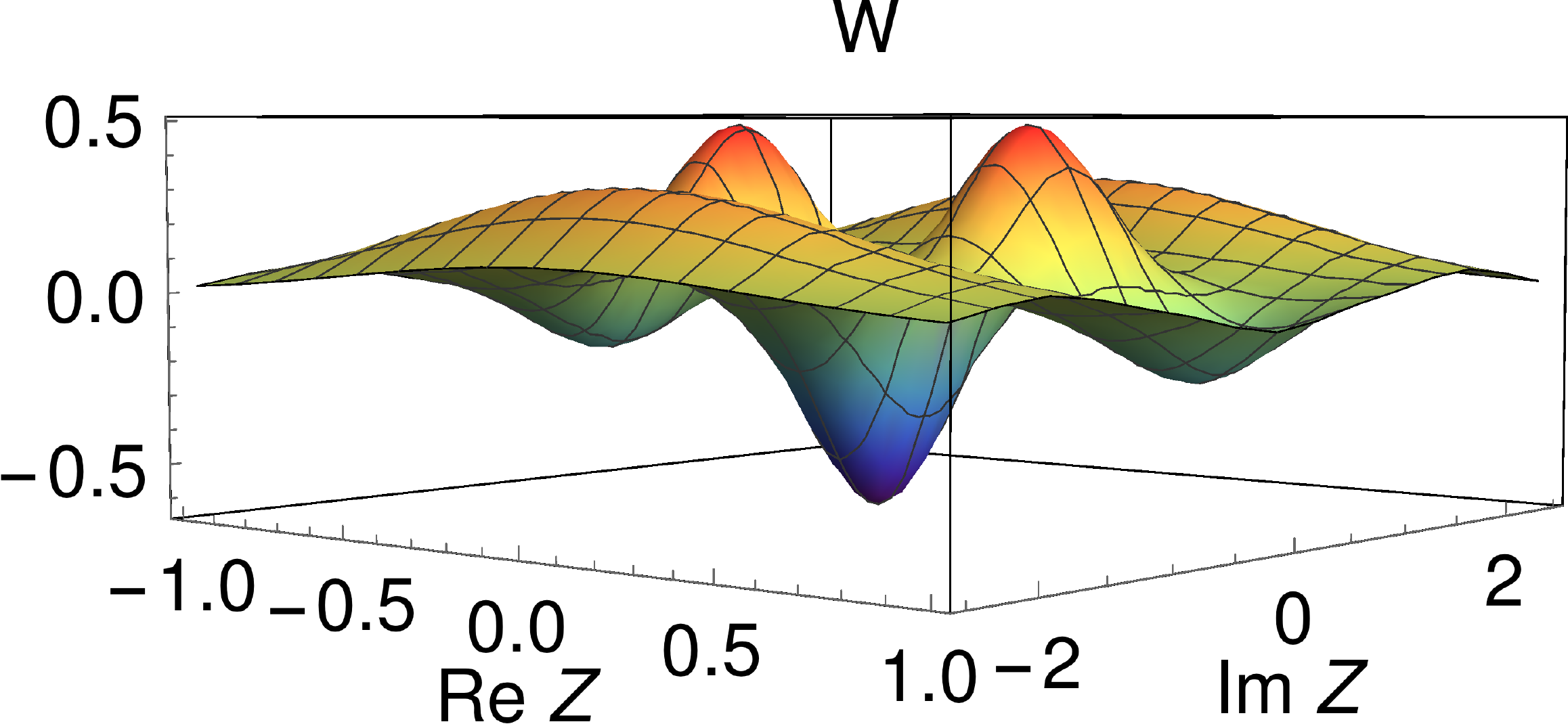}
  \label{fig5c}}
\subfigure[]{
  \includegraphics[width=7.5cm,height=5cm]{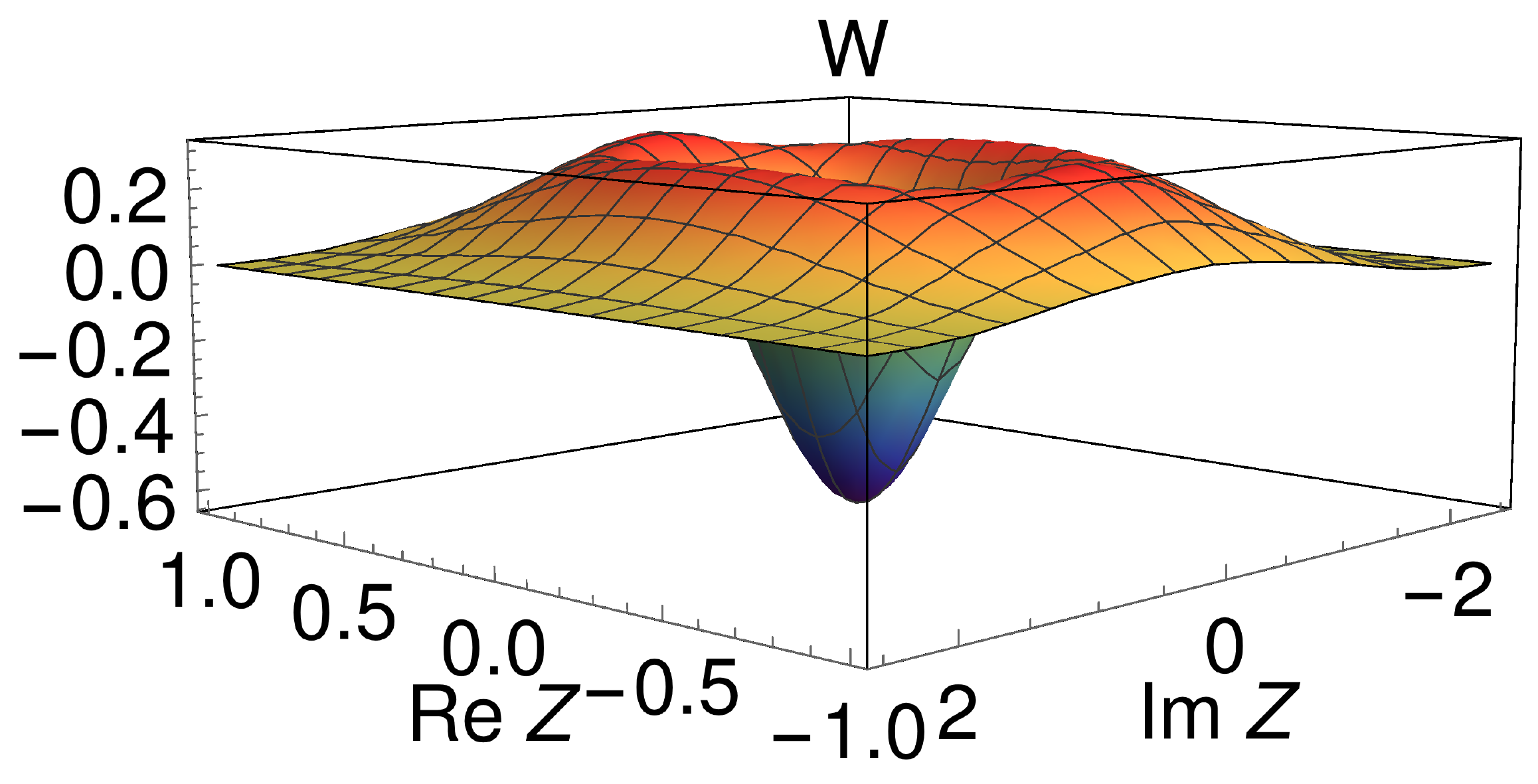}
  \label{fig5d}}
 \caption{(Color online) Wigner distribution function for generalized PSSVS \subref{fig5a} $m=14$ \subref{fig5b} $m=1$ for the even case, and those for the odd case are in \subref{fig5c}  $m=10$ \subref{fig5d} $m=1$. In all the plots we have taken $r=4, \theta=0, \lambda=\kappa=1.5$.}
  \label{WF}
\end{figure*}
compute $N_{\text{s}}^{\text{e/o}}={}_{\text{e/o}}\langle (\Delta \hat{N})^2\rangle_{\text{e/o}} -{}_{\text{e/o}}\langle \hat{N}\rangle_{\text{e/o}}$ by using the following two expressions
\begin{eqnarray}
&& _{\text{e/o}}\langle N\rangle_{\text{e/o}}=\left\langle\left[A^\dagger A+\frac{(\lambda+\kappa)^2}{4}\right]^{\frac{1}{2}}\right\rangle_{\text{e/o}}-\frac{\lambda+\kappa}{2}, \label{NumAv}\\
&& _{\text{e/o}}\langle N^2\rangle_{\text{e/o}}={}_{\text{e/o}}\langle A^\dagger A\rangle_{\text{e/o}}-(\lambda+\kappa){}_{\text{e/o}}\langle N\rangle_{\text{e/o}}. \label{NumSqAv}
\end{eqnarray}
It is straightforward to calculate the expressions in \myref{NumAv} and \myref{NumSqAv} by using \myref{AAdag1} and \myref{AAdag2}, and the results are shown in Fig.\,\ref{NS_even} and \ref{NS_odd} for even and odd PSSVS, respectively. In Fig.\,\ref{fig1a} and \ref{fig1c}, we plot the variation of $N_\text{s}^{\text{e}}$ with respect to $r$ as a function $m$ for generalized and harmonic oscillator PSSVS, respectively. Fig.\,\ref{fig1a} demonstrates negative values of $N_\text{s}^{\text{e}}$, whereas no negative region is visible in Fig.\,\ref{fig1c}, which indicates that the photon distribution for generalized PSSVS is squeezed. A similar thing happens in Fig.\,\ref{fig1b} and \ref{fig1d} also, where we plot the variation of $N_\text{s}^{\text{e}}$ with respect to $m$. Overall, we observe that the photon number distribution is always squeezed in generalized even PSSVS (provided that we choose the $r$ and $m$ values properly), and corresponding states are nonclassical, but the PSSVS for corresponding to harmonic oscillator are not nonclassical. In case of odd PSSVS, we obtain even better results, which we show in Fig.\,\ref{NS_odd}. Here, we do not see any positive value of $N_\text{s}^{\text{o}}$ for generalized PSSVS, whereas, the $N_\text{s}^{\text{o}}$ for harmonic oscillator PSSVS are always positive. 
\subsection{Negativity of the Wigner distribution function}\label{sec33}
In order to obtain the Wigner distribution function in our case, we shall incorporate the expressions of generalized PSSVS $|\zeta,f,m\rangle_{\text{e/o}}$ from \myref{EPSSVS} and \myref{OPSSVS} in its usual form \cite{Wigner,Zelaya_Dey_Hussin}  
\begin{eqnarray} \label{WignerF}
W(z)_{\text{e/o}}=e^{2|z|^2} && \int \frac{\text{d}^2\beta}{\pi^2}\langle -\beta|\zeta,f,m\rangle_\text{e/o}~{_\text{e/o}}\langle \zeta,f,m|\beta\rangle \notag\\
&& \qquad\qquad\times 2 e^{2(\beta^\ast z-\beta z^\ast)},
\end{eqnarray} 
where $z,\beta$ are the eigenvalues of the Glauber coherent states. As we stated earlier that our aim is to quantify the nonclassicality of the generalized PSSVS with respect to the Glauber coherent states, therefore, we must take the inner products between the generalized PSSVS $|\zeta,f,m\rangle_{\text{e/o}}$ and the Glauber coherent states $|\beta\rangle$ in \myref{WignerF}. While we utilize the exact form of the even and odd PSSVS from \myref{EPSSVS} and \myref{OPSSVS} in \myref{WignerF} and introduce a change of variable $\gamma=2z$, we obtain
\begin{eqnarray}
W(\gamma)_{\text{e/o}} &=& e^{|\gamma|^2/2}\displaystyle\sum_{l_1,l_2=0}^\infty\mathcal{C}_{l_1,l_2}^{\text{e/o}}\mathcal{F}_{l_1,l_2}^{\text{e/o}}(\gamma),
\end{eqnarray}
with
\begin{alignat}{1}
& \mathcal{C}_{l_1,l_2}^{\text{e}} = \frac{2(-\tanh r/2)^{k_1+k_2}e^{i(l_1-l_2)\theta}(2k_1)!(2k_2)!}{\pi\mathcal{N}^{\text{e}^2}_{\xi,f,m} k_1!k_2!\sqrt{(2l_1)!(2l_2)!}f(2l_1)!f(2l_2)!}, \label{Function2}\\
& \mathcal{C}_{l_1,l_2}^{\text{o}} = \frac{2\mathcal{N}^{\text{o}^{-2}}_{\xi,f,m}(-\tanh r/2)^{k'_1+k'_2}e^{i(l_1-l_2)\theta}(2k'_1)!(2k'_2)!}{\pi k'_1!k'_2!\sqrt{(2l_1+1)!(2l_2+1)!}f(2l_1+1)!f(2l_2+1)!}, \label{Function3} \\
& \mathcal{F}_{l_1,l_2}^{\text{e}}(\gamma) = \frac{(-1)^{2(l_1+l_2)}}{\sqrt{(2l_1)!(2l_2)!}}\frac{\partial^{2(l_1+l_2)}}{\partial\gamma^{2l_1}\partial\gamma^{\ast2l_2}}e^{-|\gamma|^2}. \label{Function}
\end{alignat}
Here, we have used the identity $\int\frac{\text{d}^2\beta}{\pi}e^{-|\beta|^2}e^{\gamma\beta^\ast-\gamma^\ast\beta}=e^{-|\gamma|^2}$ and parameterized $k_1=m+l_1, k_2=m+l_2, k'_1=k_1+1, k'_2=k_2+1$. In order to write \myref{Function} in a more compact form, we shall utilize the fact that the derivative of any analytic function $f(z,z^\ast)$ with respect to $z$ is independent of $z^\ast$ and vice-versa, as well as employ the Rodrigues formula for the associated Laguerre polynomials to obtain
\begin{eqnarray} \label{Function1}
&& \mathcal{F}_{l_1,l_2}^{\text{e}}(\gamma)= \\
&&\left\{ \begin{array}{ll}
\sqrt{\frac{(2l_1)!}{(2l_2)!}}e^{-4|z|^2}(2z)^{2(l_2-l_1)} L_{2l_1}^{2(l_2-l_1)}(4|z|^2),
& l_2\geq l_1 \\ \sqrt{\frac{(2l_2)!}{(2l_1)!}}e^{-4|z|^2}(2z^\ast)^{2(l_1-l_2)} L_{2l_2}^{2(l_1-l_2)}(4|z|^2), & l_2\leq l_1, \end{array}\right. \notag
\end{eqnarray} 
where we have have restored the original variable by identifying $z=\gamma/2$. The function $\mathcal{F}_{l_1,l_2}^\text{o}(\gamma)$ corresponding to the odd case is obtained by replacing all $2l_1$ and $2l_2$ in both \myref{Function} and \myref{Function1} by $2l_1+1$ and $2l_2+1$, respectively. The behavior of the Wigner function for both the even and odd generalized PSSVS are shown in Fig.\,\ref{WF}. In both of the cases, we find that the negativity in Wigner function becomes stronger while we subtract more photons from the squeezed vacuum states. For instance, in Fig.\,\ref{fig5a} we have taken $m=14$ which implies that we have subtracted 28 photons, whereas in Fig.\,\ref{fig5b} $m=1$, that means only 2 photons are subtracted, and it is clear that the nonclassicality in the former case is higher than the latter. A similar type of observation can also be made for the odd cases in Fig.\,\ref{fig5c} and \ref{fig5d}. In these cases, although we do not notice an increase of negativity of the main peak, but we see more negative peaks around the principal peak when higher number of photons are subtracted. We have no quantitative analysis in our hand though, with which we can claim that the odd PSSVS have a stronger nonclassicality with more number of photons being subtracted, however, intuitively it is more likely. Nevertheless, it is always true that the PSSVS are more nonclassical than the squeezed vacuum states, which is true in both the usual and generalized cases. The notion of higher degree of nonclassicality for the generalized PSSVS than those of the harmonic oscillator also sustains in this method also, which we do not present here. 
\section{Concluding remarks}\label{sec5}
We have proposed a method for the generalization of the PSSVS along with an example of the trigonometric P\"oschl-Teller potential on which our general framework has been applied. We explore three different methods; such as, quadrature squeezing, number squeezing and negativity of the Wigner function to show the nonclassical properties of the PSSVS for the P\"oschl-Teller model. A part of the discussion consists of the generalization of the above three schemes which become suitable for the analysis of nonclassicality of the generalized PSSVS. We find that the degree of nonclassicality can be enhanced by subtracting more photons from the generalized squeezed vacuum states. Therefore, utilization of the generalized PSSVS may be more advantageous over the squeezed vacuum states. We also compare our results with those emerging from the harmonic oscillator limits, and we observe that generalized PSSVS often demonstrate more nonclassicality compared to the harmonic oscillator cases. Thus, the generalized PSSVS provide a twofold enhancement of the nonclassicality with respect to the harmonic oscillator squeezed vacuum states and we believe that such states will bring fascinating outcomes for the study of quantum information.

So far, all results are theoretical, however, it is interesting to notice that the states can be constructed analytically and most of the analysis of nonclassicality can also be carried out analytically. A proper understanding of the states in the laboratory can be achieved only after the realization of some basic nonlinear quantum optical states in real life, which is currently under intense investigation. For the time being we believe that a lot of theoretical research are yet to be performed for a deeper understanding of the framework. As an immediate follow up, it would be worth studying the given protocol for some other general models and verify whether our results hold in those cases also, which may effectively shed some light on the given direction.  
\section{Acknowledgements} S.\,D.\,acknowledges the support of research grant (DST/INSPIRE/04/2016/001391) from DST, Govt.\,of India. S.\,S.\,N.\,is supported by a project assistant research fellowship by DST, Govt.\,of India.

\begin{thebibliography}{99}	

\bibitem{Eisert_Scheel_Plenio}
J.\,Eisert, S.\,Scheel and M.\,B.\,Plenio,
\newblock Distilling Gaussian states with Gaussian operations is impossible,
\newblock {Phys.\,Rev.\,Lett.}\,\textbf{89}, 137903 (2002).

\bibitem{Bartlett_Sanders}
S.\,D.\,Bartlett and B.\,C.\,Sanders,
\newblock Universal continuous-variable quantum computation: requirement of optical nonlinearity for photon counting,
\newblock {Phys.\,Rev.\,A} \textbf{65}, 042304 (2002).

\bibitem{Wenger_Tualle-Brouri_Grangier}
J.\,Wenger, R.\,Tualle-Brouri and P.\,Grangier,
\newblock Non-Gaussian statistics from individual pulses of squeezed light,
\newblock {Phys.\,Rev.\,Lett.}\,\textbf{92}, 153601 (2004).

\bibitem{Lvovsky_etal}
A.\,I.\,Lvovsky~et al.,
\newblock Quantum state reconstruction of the single-photon Fock state,
\newblock {Phys.\,Rev.\,Lett.}\,\textbf{87}, 050402 (2001).

\bibitem{Chen_etal1}
Z.\,Chen, Y.\,Zhou and J.-T.\,Shen,
\newblock Photon antibunching and bunching in a ring-resonator waveguide quantum electrodynamics system,
\newblock {Opt.\,Lett.}\,\textbf{41}, 3313-3316 (2016).

\bibitem{Chen_etal2}
Z.\,Chen, Y.\,Zhou and J.-T.\,Shen,
\newblock Dissipation-induced photonic-correlation transition in waveguide-QED systems,
\newblock {Phys.\,Rev.\,A} \textbf{96}, 053805 (2017).

\bibitem{Chen_etal}
Z.\,Chen, Y.\,Zhou and J.-T.\,Shen,
\newblock Correlation signatures for a coherent three-photon scattering in waveguide quantum electrodynamics,
\newblock {Opt.\,Lett.}\,\textbf{45}, 2559-2562 (2020).

\bibitem{Ourjoumtsev_Jeong_Tualle-Brouri_Grangier}
A.\,Ourjoumtsev, H.\,Jeong, R.\,Tualle-Brouri and P.\,Grangier,
\newblock Generation of optical {‘Schr{\"o}dinger cats’ from} photon number states,
\newblock {Nature} \textbf{448}, 784 (2007).

\bibitem{Wu_Kimble_Hall_Wu}
L.-A.\,Wu, H.\,J.\,Kimble, J.\,L.\,Hall and H.\,Wu,
\newblock Generation of squeezed states by parametric down conversion,
\newblock {Phys.\,Rev.\,Lett.}\,\textbf{57}, 2520 (1986).

\bibitem{Wakui_Takahashi_Furusawa_Sasaki}
K.\,Wakui, H.\,Takahashi, A.\,Furusawa and M.\,Sasaki,
\newblock Photon subtracted squeezed states generated with periodically poled {KTiOPO$_4$},
\newblock {Opt.\,Express} \textbf{15}, 3568--3574 (2007).

\bibitem{Dodonov}
V.\,V.\,Dodonov,
\newblock Nonclassical states in quantum optics: a squeezed review of the first 75 years,
\newblock {J.\,Opt.\,B: Quantum Semiclas.\,Opt.}\,\textbf{4}, R1 (2002).

\bibitem{Dey_Fring_Hussin_Review}
S.\,Dey, A.\,Fring and V.\,Hussin,
\newblock A squeezed review on coherent states and nonclassicality for {non-Hermitian systems with minimal length},
\newblock In {\em Coherent states and their applications}, Springer Proc.\,Phys.\,\textbf{205}, 209--242 (2018).

\bibitem{Manko_Marmo_Sudarshan_Zaccaria}
V.\,I.\,Man'ko, G.\,Marmo, E.\,C.\,G.\,Sudarshan and F.\,Zaccaria,
\newblock $f$-oscillators and nonlinear coherent states,
\newblock {Phys.\,Scr.}\,\textbf{55}, 528 (1997).

\bibitem{Sivakumar}
S.\,Sivakumar,
\newblock Studies on nonlinear coherent states,
\newblock {J.\,Opt.\,B: Quantum Semiclas.\,Opt.}\,\textbf{2}, R61 (2000).

\bibitem{Filho_Vogel}
R.\,L.\,M.\,Filho and W.\,Vogel,
\newblock Nonlinear coherent states,
\newblock {Phys.\,Rev.\,A} \textbf{54}, 4560 (1996).

\bibitem{Biedenharn}
L.\,C.\,Biedenharn,
\newblock The quantum group {$SU_q(2)$ and} a $q$-analogue of the boson operators,
\newblock {J.\,Phys.\,A: Math.\,Gen.}\,\textbf{22}, L873 (1989).

\bibitem{Macfarlane}
A.\,J.\,Macfarlane,
\newblock On $q$-analogues of the quantum harmonic oscillator and the quantum group {$SU(2)_q$},
\newblock {J.\,Phys.\,A: Math.\,Gen.}\,\textbf{22}, 4581 (1989).

\bibitem{Dey_Hussin_atom}
S.\,Dey and V.\,Hussin,
\newblock Squeezed atom laser for {Bose-Einstein} condensate with minimal length,
\newblock {Int.\,J.\,Theor.\,Phys.}\,\textbf{58}, 3138--3148 (2019).

\bibitem{Dey}
S.\,Dey,
\newblock $q$-deformed noncommutative cat states and their nonclassical properties,
\newblock {Phys.\,Rev.\,D} \textbf{91}, 044024 (2015).

\bibitem{Dey_Hussin_Photon}
S.\,Dey and V.\,Hussin,
\newblock Noncommutative $q$-photon-added coherent states,
\newblock {Phys.\,Rev.\,A} \textbf{93}, 053824 (2016).

\bibitem{Zelaya_Dey_Hussin}
K.\,Zelaya, S.\,Dey and V.\,Hussin,
\newblock Generalized squeezed states,
\newblock {Phys.\,Lett.\,A} \textbf{382}, 3369--3375 (2018).

\bibitem{Kim_Son_Buzek_Knight}
M.\,S.\,Kim, W.\,Son, V.\,Bu{\v{z}}ek and P.\,L.\,Knight,
\newblock Entanglement by a beam splitter: Nonclassicality as a prerequisite for entanglement,
\newblock {Phys.\,Rev.\,A} \textbf{65}, 032323 (2002).

\bibitem{Dey_Hussin}
S.\,Dey and V.\,Hussin,
\newblock Entangled squeezed states in noncommutative spaces with minimal length uncertainty relations,
\newblock {Phys.\,Rev.\,D} \textbf{91}, 124017 (2015).

\bibitem{Dey_Fring_Hussin}
S.\,Dey, A.\,Fring and V.\,Hussin,
\newblock Nonclassicality versus entanglement in a noncommutative space,
\newblock {Int.\,J.\,Mod.\,Phys.\,B} \textbf{31}, 1650248 (2017).

\bibitem{Wang_Goorskey_Xiao}
H.\,Wang, D.\,Goorskey and M.\,Xiao,
\newblock Enhanced {Kerr} nonlinearity via atomic coherence in a three-level atomic system,
\newblock {Phys.\,Rev.\,Lett.}\,\textbf{87}, 073601 (2001).

\bibitem{Gambetta_etal}
A.\,Gambetta et al.,
\newblock Real-time observation of nonlinear coherent phonon dynamics in single-walled carbon nanotubes,
\newblock {Nature Phys.}\,\textbf{2}, 515 (2006).

\bibitem{Yan_Zhu_Li}
Y.\,Yan, J.-P.\,Zhu and G.-X.\,Li,
\newblock Preparation of a nonlinear coherent state of the mechanical resonator in an optomechanical microcavity,
\newblock {Opt.\,Express} \textbf{24}, 13590--13609 (2016).

\bibitem{Agarwal_Biswas_1}
A.\,Biswas and G.\,S.\,Agarwal,
\newblock Nonclassicality and decoherence of photon-subtracted squeezed states,
\newblock {Phys.\,Rev.\,A} \textbf{75}, 032104 (2007).

\bibitem{Gerry_Knight_Book}
C.\,Gerry and P.\,Knight,
\newblock {\em Introductory quantum optics},
\newblock Cambridge Univ.\,Press: New York (2005).

\bibitem{Trifonov}
D.\,A.\,Trifonov,
\newblock Generalized uncertainty relations and coherent and squeezed states,
\newblock {J.\,Opt.\,Soc.\,Am.\,A} \textbf{17}, 2486--2495 (2000).

\bibitem{Quesne_Penson_Thachuk}
C.\,Quesne, K.\,A.\,Penson and V.\,M.\,Tkachuk,
\newblock Maths-type $q$-deformed coherent states for $q>1$,
\newblock {Phys.\,Lett.\,A} \textbf{313}, 29--36 (2003).

\bibitem{Kwek_Kiang}
L.\,C.\,Kwek and D.\,Kiang,
\newblock Nonlinear squeezed states,
\newblock {J.\,Phys.\,B: Quantum Semiclas.\,Opt.}\,\textbf{5}, 383 (2003).

\bibitem{Naderi_Soltanolkotabi_Roknizadeh}
M.\,H.\,Naderi, M.\,Soltanolkotabi and R.\,Roknizadeh,
\newblock New photon-added and photon-depleted coherent states associated with inverse $q$-boson operators: nonclassical properties,
\newblock {J.\,Phys.\,A: Math.\,Gen.}\,\textbf{37}, 3225 (2004).

\bibitem{Ching_Ng}
C.\,L.\,Ching and W.\,K.\,Ng,
\newblock Generalized coherent states under deformed quantum mechanics with maximum momentum,
\newblock {Phys.\,Rev.\,D} \textbf{88}, 084009 (2013).

\bibitem{Ramirez_Reboiro}
R.\,Ram{\'\i}rez and M.\,Reboiro,
\newblock Squeezed states from a quantum deformed oscillator Hamiltonian,
\newblock {Phys.\,Lett.\,A} \textbf{380}, 1117--1124 (2016).

\bibitem{Fakhri_Hashemi}
H.\,Fakhri and A.\,Hashemi,
\newblock Nonclassical properties of the $q$-coherent and $q$-cat states of the {Biedenharn-Macfarlane} $q$ oscillator with $q>1$,
\newblock {Phys.\,Rev.\,A} \textbf{93}, 013802 (2016).

\bibitem{Noormandipour_Tavassoly}
A.\,NoormandiPour and M.\,K.\,Tavassoly,
\newblock $f$-deformed squeezed vacuum and first excited states, their superposition and corresponding nonclassical properties,
\newblock {Commun.\,Theor.\,Phys.}\,\textbf{61}, 521 (2014).

\bibitem{Roy_Roy}
B.\,Roy and P.\,Roy,
\newblock New nonlinear coherent states and some of their nonclassical properties,
\newblock {J.\,Opt.\,B: Quantum Semiclas.\,Opt.}\,\textbf{2}, 65 (2000).

\bibitem{Antoine_Gazeau_Monceau_Klauder_Penson}
J.\,P.\,Antoine, J.\,P.\,Gazeau, P.\,Monceau, J.\,R.\,Klauder and K.\,A.\,Penson,
\newblock Temporally stable coherent states for infinite well and {P{\"o}schl--Teller} potentials,
\newblock {J.\,Math.\,Phys.}\,\textbf{42}, 2349--2387 (2001).

\bibitem{Reed}
M.\,Reed and B.\,Simon,
\newblock {\em Methods of modern mathematical physics},
\newblock {Academic press: New York} (1978).

\bibitem{Dey_Fring_squeezed}
S.\,Dey and A.\,Fring,
\newblock Squeezed coherent states for noncommutative spaces with minimal length uncertainty relations,
\newblock {Phys.\,Rev.\,D} \textbf{86}, 064038 (2012).

\bibitem{Dey_Fring_Gouba_Castro}
S.\,Dey, A.\,Fring, L.\,Gouba and P.\,G.\,Castro,
\newblock Time-dependent $q$-deformed coherent states for generalized uncertainty relations,
\newblock {Phys.\,Rev.\,D} \textbf{87}, 084033 (2013).

\bibitem{Mandel}
L.\,Mandel,
\newblock {Sub-Poissonian} photon statistics in resonance fluorescence,
\newblock {Opt.\,Lett.}\,\textbf{4}, 205--207 (1979).

\bibitem{Zelaya}
K.\,Zelaya, O.\,Rosas-Ortiz, Z.\,Blanco-Garcia and S.\,Cruz y Cruz,
\newblock Completeness and nonclassicality of coherent states for generalized oscillator algebras,
\newblock {Adv.\,Math.\,Phys.}\,\textbf{2017}, 7168592 (2017).

\bibitem{Wigner}
E.\,Wigner,
\newblock On the quantum correction for thermodynamic equilibrium,
\newblock {Phys.\,Rev.}\,\textbf{40}, 749 (1932).

\end{thebibliography}


\end{document}